\documentclass[twocolumn]{aastex701}

\usepackage{amsmath}
\usepackage{graphicx}  
\usepackage{tikz}      
\usetikzlibrary{positioning, decorations.pathreplacing} 
\usepackage{subcaption}
\usepackage{adjustbox}
\usepackage{comment}
\usepackage{CJKutf8}
\usepackage{hyperref}

\newcommand{\affilcaltech}{\affiliation{Department of Physics, California Institute of Technology, 1200 E. California Boulevard, Pasadena, CA 91125, USA}}
\newcommand{\affiljpl}{\affiliation{Jet Propulsion Laboratory, 
California Institute of Technology, 
4800 Oak Grove Drive, 
Pasadena, CA 91109, USA}}
\newcommand{\affilkasi}{\affiliation{Korea Astronomy and Space Science Institute (KASI), 776 Daedeok-daero, Yuseong-gu, Daejeon 34055, Republic of Korea}}
\newcommand{\affilipac}{\affiliation{IPAC, California Institute of Technology, 770 S. Wilson Ave, Pasadena, CA 91125, USA}}

\begin{document}
\begin{CJK*}{UTF8}{gbsn}

\title{Simulating Spectral Confusion in SPHEREx Photometry and Redshifts}

\author[0009-0009-1219-5128]{Zhaoyu Huai}
\email[show]{zhuai@caltech.edu}
\affilcaltech

\author[0000-0002-5710-5212]{James J. Bock}
\email{jjb@astro.caltech.edu}
\affilcaltech
\affiljpl

\author[0000-0002-5437-0504]{Yun-Ting Cheng \begin{CJK*}{UTF8}{bsmi}(鄭昀庭)\end{CJK*}}
\email{ycheng3@caltech.edu}
\affilcaltech
\affiljpl

\author[0009-0008-8066-446X]{Jean Choppin de Janvry}
\email{jean.choppindejanvrydev@gmail.com}
\affiljpl
\affiliation{Berkeley Center for Cosmological Physics, University of California, Berkeley, CA 94720, USA}
\affiliation{Lawrence Berkeley National Laboratory, Berkeley, California 94720, USA}

\author[0000-0002-6503-5218]{Sean Bruton}
\email{sbruton@caltech.edu}
\affilcaltech

\author[0000-0002-1630-7854]{James R. Cheshire IV}
\email{cheshire@caltech.edu}
\affilcaltech

\author[0000-0002-4650-8518]{Brendan P. Crill}
\email{brendan.p.crill@jpl.nasa.gov}
\affiljpl

\author[0000-0002-0867-2536]{Olivier Dor\'{e}}
\email{olivier.p.dore@jpl.nasa.gov}
\affiljpl
\affilcaltech

\author[0000-0002-3745-2882]{Spencer W. Everett}
\email{severett@caltech.edu}
\affilcaltech

\author[0000-0002-9382-9832]{Andreas L. Faisst}
\email{afaisst@caltech.edu}
\affilipac

\author[0000-0002-9330-8738]{Richard M. Feder}
\email{rmfeder@berkeley.edu}%
\affiliation{Berkeley Center for Cosmological Physics, University of California, Berkeley, CA 94720, USA}
\affiliation{Lawrence Berkeley National Laboratory, Berkeley, California 94720, USA}

\author[0000-0002-2770-808X]{Woong-Seob Jeong}
\email{jeongws@kasi.re.kr}
\affilkasi

\author[0000-0003-1647-3286]{Yongjung Kim}
\email{yjkim.ast@gmail.com}
\affiliation{School of Liberal Studies, Sejong University, 209 Neungdong-ro, Gwangjin-Gu, Seoul 05006, Republic of Korea}
\affiliation{Department of Physics and Astronomy, Sejong University, 209 Neungdong-ro, Gwangjin-Gu, Seoul 05006, Republic of Korea}

\author[0000-0003-1954-5046]{Bomee Lee}
\email{bomee@kasi.re.kr}
\affilkasi

\author[0000-0001-5382-6138]{Daniel C. Masters}
\email{dmasters@ipac.caltech.edu}
\affilcaltech
\affilipac

\begin{abstract}

We model the impact of source confusion on photometry and the resulting spectrophotometric redshifts for SPHEREx, a NASA Medium-Class Explorer that is carrying out an all-sky near-infrared spectral survey. Spectral confusion from untargeted background galaxies degrades sensitivity and introduces a spectral bias. Using interpolated spectral energy distributions (SEDs) from the COSMOS2020 catalog, we construct a Monte Carlo library of confusion spectra that captures the cumulative impact from faint galaxies. By injecting confusion realizations into galaxy SEDs and performing forced photometry at known source positions, we quantify photometric and redshift error and bias. For our current expected selection of sources for the cosmology analysis, we find typical 1-$\sigma$ confusion levels range from $0.8-3.8\ \mu\mathrm{Jy}$ across $0.75-5.0\ \mu\mathrm{m}$. While negligible at full-sky survey depth, spectral confusion becomes significant in the SPHEREx deep fields, reducing the number of intermediate-precision redshifts and inducing a small systematic overestimation in redshift. In parallel, we also model targeted source blending from beam overlaps, which contributes additional photometric noise without systematic redshift bias, provided that positions are known exactly. Together, confusion and blending vary with the depth of the selected reference sample, revealing a trade-off, where deeper selections reduce confusion but increase blending-induced noise. Our methodology informs optimization of the SPHEREx deep-field selection strategy and future treatments of stellar source blending and confusion.

\end{abstract}


\section{Introduction} \label{sec:intro}

    The Spectro-Photometer for the History of the Universe, Epoch of Reionization, and Ices Explorer (SPHEREx, \citealt{Dore2015}) is a NASA astrophysics mission that launched in March 2025. Over its two-year baseline mission, SPHEREx will conduct four full-sky surveys by repeatedly imaging the sky in a coordinated scan strategy that builds up full coverage across 102 wavelength bands over $0.75-5\ \mu\mathrm{m}$ \citep{phil18, brendan20}.

    SPHEREx is designed to constrain the physics of inflation, probe galaxy evolution, and explore the origin of water in planetary systems \citep{Dore2015, dore16scienceI, dore18scienceII}. Signatures of inflation, such as primordial non-Gaussianity, leave subtle imprints on the large-scale structure (LSS) at later times \citep{olivier08, alvarez14fnl}. Constraining this feature requires accurate 3D galaxy maps, which relies on sufficient galaxy number density and redshift accuracy to trace the underlying matter field. \cite{stickley} outlined spectrophotometric redshift estimates using template fitting, further characterized in \cite{richard24} through updated simulations. However, previous studies have not considered the impact of spectral confusion and blending. In this work, we assess how these effects influence galaxy selection and redshift accuracy, with implications for SPHEREx cosmology.
    
    SPHEREx performs forced photometry at fixed source positions specified in the SPHEREx Reference Catalog (RC, \citealt{REFCAT_PAPER}), which uses external survey catalogs for target positions (Section~\ref{sec:RC}). Photometric and redshift quality depends on the catalog selection depth, initially chosen near the instrument sensitivity limit. With that, we consider two effects: 
    \begin{enumerate}
        \item \emph{Blending} --- spatial overlap of targeted sources within the angular resolution of the SPHEREx beam, on which simultaneous forced photometry is done.
        \item \emph{Spectral confusion} --- flux contamination to targeted sources from faint, nearby galaxies not included in our photometric selection, which may be present in the SPHEREx RC but are not chosen for forced photometry.
    \end{enumerate} 
    We adopt the classical definitions and distinguish between \emph{blending} and \emph{confusion} \citep{murdoch73, condon, franceschini86, helou&beichman90, rieke95, dhogg01, legache03, takeuchi04}: \emph{blending} arises from overlapping targeted sources due to chance alignments, and \emph{confusion} is the cumulative flux from untargeted sources. Our RC-based approach extends this boundary: some untargeted sources may be known, but are excluded by the catalog selection for example. 

    Confusion and blending have long been recognized as limiting factors in deep extragalactic surveys. 
    Classical confusion is severe in infrared, submillimeter, and radio single-dish telescopes, where limited angular resolution leads to flux contamination from unresolved sources \citep{helou&beichman90, ermash20, dole2004, frayer06}. 
    Extensive confusion studies at the image level from the \textit{Spitzer Space Telescope} \citep{werner04spitzer} have shown that at $24\ \mu\mathrm{m}$, $\sim\!70\%$ of the background was resolved \citep{dole04a, papovish04}, but confusion limits become dominant at longer wavelengths. 
    The Herschel  Multi-tiered Extragalactic Survey (HerMES; \citeauthor{pilbratt10herschel} \citeyear{pilbratt10herschel}; \citeauthor{hermesOliver} \citeyear{hermesOliver}) used stacking to disentangle confusing sources \citep{viero13}. 
    The proposed \textit{PRobe far-Infrared Mission for Astrophysics} (PRIMA; \citeauthor{moullet23prima} \citeyear{moullet23prima}) explored polarization for confusion mitigation \citep{donnellan24, bethermin24}.

    Blending has also been a challenge in galaxy surveys. For the \textit{Rubin Observatory} Legacy Survey of Space and Time (LSST; \citeauthor{lsst} \citeyear{lsst}), \cite{sanchez21} expects $\sim\!62\%$ of detected galaxies to experience $>1\%$ flux contamination from overlapping sources, correlating with degraded photometric redshift performance \citep{lsstblending, BPZ00, flexboost}. Other surveys experience similar challenges, particularly in weak lensing measurements \citep{blendingreview, nourbakhsh22, euclidshear}.

    Prior work largely treats confusion and blending separately at the source detection or image level, without considering their impact in forced photometry with a coupled reference catalog of sources, and the downstream effects on galaxy redshifts. In this work, we extend these efforts into the spectrophotometric redshift regime of SPHEREx. Spectral confusion can systematically bias redshifts, while blending primarily increases photometric noise without inherently biasing redshifts, assuming accurate astrometry. Because SPHEREx performs forced photometry on known, targeted source positions, the two effects are coupled: both depend on the selection depth of the catalog and the spatial clustering. We use the SPHEREx Sky Simulator \citep{Crill25} combined with the deep COSMOS survey \citep{COSMOS} to quantify how both blending and spectral confusion affect photometric and redshift accuracy. 

    This paper is organized as follows: Section~\ref{sec2} introduces the SPHEREx survey setup, photometry tool, reference catalog, and redshift estimates. Section~\ref{sec3} describes the method for modeling the blending effects, while Section~\ref{sec:lib_method} details the simulation of spectral confusion. Redshift results are presented in Section~\ref{sec5}, highlighting the trade-off between blending and confusion. We discuss broader implications in Section~\ref{sec:discussion} and conclude in Section~\ref{sec:conclusion}. In this work, we use the AB-magnitude system \citep{abmag}.

\section{SPHEREx Catalog, Photometry, and Redshift Framework}
\label{sec2}
    
    \subsection{SPHEREx Overview}

        SPHEREx will conduct an all-sky spectroscopic survey in 102 near-infrared channels spanning $0.75\ \mu\mathrm{m}$ to $5\ \mu\mathrm{m}$. It employs a 20 cm telescope with two 1$\times$3 mosaics of H2RG detector arrays \citep{Chi2025}, using linear variable filters (LVFs) mounted above the detectors to deliver spectral resolving powers of $R = \lambda/\Delta\lambda = 35-130$ \citep{phil18}. Each detector mosaic covers a $3.5^{\circ} \times 11.3^{\circ}$ field of view at $6\farcs15$ pixel resolution, enabling a high throughput system for efficient full-sky coverage. 

        With a sun-synchronous polar orbit, SPHEREx will obtain at least four independent observations per wavelength channel for each target over the full sky in its two-year baseline mission. It will collect significantly deeper observations in the North and South Ecliptic poles (NEP \& SEP) \citep{Dore2015, brendan20}, measuring the full spectral coverage $\sim\!100$ times per source over these 100 $\mathrm{deg}^2$ each \citep{richard24}. In these regions, the number of repeated observations increases with ecliptic latitude, reaching up to four hundred measurements at the center of the field. While this enhanced depth offers huge gains in sensitivity and redshift accuracy, it also introduces complexity due to spatially varying sensitivity and a higher fractional contribution from spectral confusion when exploiting its full depth. For clarity, we refer to the two coverage depths as ``full-sky" and ``deep-field" throughout this work.

    \subsection{SPHEREx Photometry}

        The SPHEREx pipeline utilizes the tool Tractor \citep{Tractor16softward} to perform forced photometry at known, fixed source positions \citep{nyland17} from external catalogs. Given a point spread function (PSF), Tractor forward models the aggregate flux from all sources in the scene and optimizes the flux of each target to maximize the likelihood of matching the imaging data, enabling robust flux measurements even in crowded fields.

        In the SPHEREx survey, each source is observed at a unique set of wavelengths determined by its position on the detectors in each exposure and the wavelength-dependent response of LVFs. These observations form the source's \textit{primary} photometry, which is then binned into 102-band \textit{secondary} photometry. This binning typically includes up to 400 observations per spectral channel in the deep fields and $\sim\!4$ observations elsewhere. A simulated spectrum demonstrating the spectral coverage is shown in Figure 4 of \citet{Crill25} (also see \citeauthor{EdwardZhang25} \citeyear{EdwardZhang25}).

        To model the expected data products, we use the SPHEREx Sky Simulator \citep{Crill25}. The Simulator generates realistic scenes as seen by SPHEREx from input catalogs by simulating all major instrumental noise sources, including lab-measured detector read noise and dark current. It also incorporates dominant astrophysical backgrounds, primarily zodiacal light (ZL) - scattering and thermal emission from interplanetary dust \citep{zodi1, zodi2}. The Simulator reproduces optical effects by convolving source fluxes with SPHEREx’s wavelength-dependent PSFs \citep{psfSymons}, which broadens from $\mathrm{FWHM}\sim4\farcs5$ (at $0.75\ \mu\mathrm{m}$) to $\sim7^{\prime\prime}$ (at $5\ \mu\mathrm{m}$) due to diffraction, along with other position-dependent optical aberrations across the detector plane. Simulated scenes are initially constructed at high resolution, using a default upsampling factor of 5 to $\sim\!1\farcs23$, to enable more accurate source placement. These images are then downsampled to the native SPHEREx pixel scale of $6\farcs15$, with noise components added at this stage. The Simulator can output either images or catalog-level photometry.

        \subsubsection{SPHEREx Reference Catalog}
        \label{sec:RC}

            SPHEREx performs forced photometry on sources from the SPHEREx Reference Catalog (RC), which integrates data from the upcoming Rubin/LSST \citep{lsst}, DESI Legacy Imaging Survey (LS; \citeauthor{DeyDESI} \citeyear{DeyDESI}), Gaia \citep{gaia}, the Two Micron All Sky Survey (2MASS; \citeauthor{2mass} \citeyear{2mass}), Pan-STARRS1 \citep{PS_catalog}, AllWISE \citep{WrightWISE}, and CatWISE 2020 \citep{CatWISE}, covering the full sky. The RC provides source positions, broadband photometry, and flags for all sources, while morphology and spectroscopic redshifts are included where available. A preliminary selection cut is applied to determine which sources will be targeted for SPHEREx photometry and redshift estimation. This cut balances instrument sensitivity against confusion noise. While a deeper and more inclusive threshold may increase completeness at higher redshifts, it also raises the risk of source blending.
            
\begin{figure}
                \centering
                \includegraphics[width=1.0\linewidth]{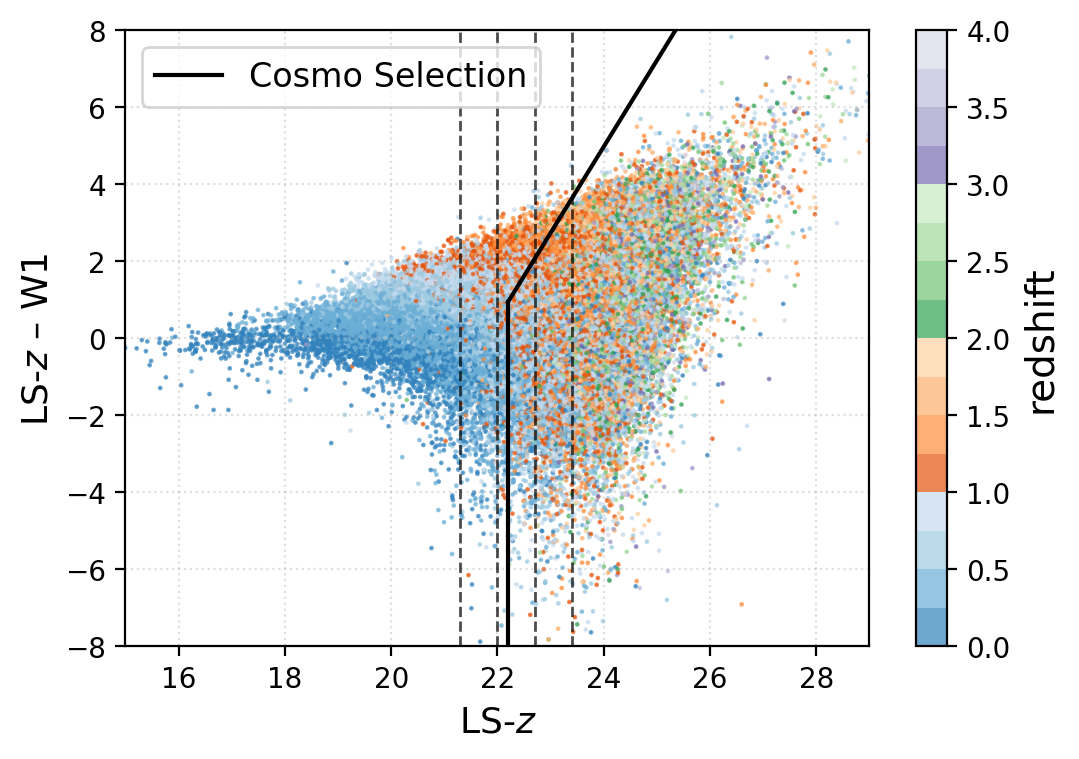}
                \caption{Color–magnitude diagram (LS-$z$ – WISE-1 vs. LS-$z$) for COSMOS and GAMA galaxies cross-matched to the LS catalog, resampled according to their survey footprints (1.27 and 217 deg$^2$, respectively). Points are color-coded by redshift. The solid black lines show the fiducial cosmology selection defined in Equation~\ref{eq:cut}. Vertical dashed lines indicate candidate constant $z$-band magnitude cuts at 21.3, 22.0, 22.7, and 23.4, which will be explored in following blending and confusion analyses.}
                \label{fig:sec2/cut}
            \end{figure}

            We apply a color-magnitude selection cut using the LS-$z$ band and WISE-1 (W1) forced photometry.
            The LS-$z$ band is less affected by Galactic extinction than bluer bands (e.g., \textit{g} or \textit{r}), helping to maintain uniformity across the sky, while W1 extends coverage into the infrared. Together, LS-$z$ and W1 span most of the SPHEREx wavelength range, providing anchored color measurements. This selection also supports uniform external tracer samples for cosmology. 
            Sources that fail this cut are excluded from forced photometry and contribute instead to spectral confusion. In this work, we define the targeted (cosmology) sample using the following color-magnitude selection,
            \begin{align}
                (\text{LS-}z - W1 &\geq 2.24 \times \text{LS-}z - 48.78) \notag \\
                \text{OR} \quad (\text{LS-}z &< 22.2)
                \label{eq:cut}
            \end{align}
            as illustrated in Figure~\ref{fig:sec2/cut}.
            The precise cosmology selection will be refined and finalized in future work. Sources not passing this selection are referred to as untargeted galaxies. Under this selection, the per-channel signal-to-noise ratio (S/N) distribution peaks near unity, with most sources above S/N $>$ 1 and a long tail toward higher S/N. Approximately 15\% of targeted galaxies have at least one neighbor within one SPHEREx pixel distance (i.e., separated by less than $6\farcs15$), and 2.3\% have a neighbor within half a pixel, corresponding to a source density of $n\sim7.0$ per arcmin$^2$. This translates to $n=0.03$ beam$^{-1}$ and 0.1 beam$^{-1}$ at $0.75\ \mu\mathrm{m}$ and $5\ \mu\mathrm{m}$, respectively. In practice, the RC source density varies significantly across the sky, and the Galactic plane is excluded due to high crowding for cosmological studies.

    \subsection{Galaxy Catalog and Simulated SEDs}
    To simulate confusion, we use the COSMOS2020 \citep{COSMOS} and GAMA \citep{GAMADriver} galaxy catalogs. \cite{richard24} selected 166,041 COSMOS galaxies ($18 < i < 25$), and 44,124 GAMA galaxies ($i<18$) with accurate redshifts and precise photometry, and constructed high-resolution SEDs for each by fitting templates from \cite{Ilbert09} and \cite{brown14} to multiband photometry from both datasets. Missing spectral regions are interpolated/extrapolated using stellar population synthesis (SPS) and dust emission models. These SEDs are then used to simulate SPHEREx synthetic photometry. While COSMOS primarily covers the majority of the faint galaxy population, we include GAMA for its brighter galaxies.

    \subsection{Photometric Redshift Estimation}

    SPHEREx estimates redshifts based on the code developed by \cite{stickley} using a template-fitting approach, which adapts the LePHARE \citep{arnouts, Ilbert06} framework to perform $\chi^2$ minimization between observed spectra and scaled templates. We use a library of 160 galaxy spectral templates, including 31 SPS-based COSMOS templates \citep{Ilbert09} and 129 empirically measured galaxies SEDs \citep{brown14}, the same set used in \citet{richard24}. To account for intrinsic variations, dust attenuation models \citep{allen76, prevot84, calzetti00, seaton79, fm86dust}, a range of extinction $E(B-V)$ values, and intergalactic medium (IGM) attenuation \citep{madau95, madau99} are applied to generate a precomputed model grid. We compute a redshift probability density function (PDF) for each source, defined as the posterior marginalized over other parameters. The estimated photometric redshift $\hat{z}$ is the PDF's expectation value, and the redshift uncertainty $\hat{\sigma}_z$ is its second moment. The current SPHEREx continuum-fitting redshift pipeline is forecast to deliver $\sigma_z<0.003(1+z)$ for $\sim\!19$ M galaxies and $\sigma_z<0.2(1+z)$ for $\sim\!810$ M galaxies over $30,000\ \mathrm{deg}^2$ \citep{richard24}, incorporating external broadband photometry. In this work, we apply this pipeline to SPHEREx-only photometry to isolate the impact of spectral confusion on redshift accuracy and uncertainty, which helps inform selection cuts and ensure robust redshift estimates in crowded and deep-field regions.

\section{Modeling Blending Effects}
\label{sec3}

We model photometric blending from the overlapping PSFs of targeted sources using a Fisher-based approach \citep{fisher} that captures the flux covariances \footnote{Covariance treatment has been implemented in a dedicated branch of the \texttt{Tractor} repository: \url{https://github.com/dstndstn/tractor/tree/variance_update}}. This approach enables unbiased forced photometry and quantifies how blending-driven errors scale with source proximity and target density.

    \subsection{Limitations of the Tractor Photometry}

        We use Tractor for forced photometry at fixed source positions, which improves deblending performance compared to blind extraction methods. Its native flux uncertainty estimates are computed under the assumption that sources are isolated. As detailed in \cite{weaverTractor}, Tractor derives flux uncertainties by summing the squared model derivatives across pixels, weighted by the inverse noise variance, as a lower bound on the uncertainty (Equation~\ref{eq:tractor_calc}). This does not explicitly include covariances with adjacent sources from PSF overlaps, which can lead to underestimated uncertainties in crowded fields. In this work, we supplement Tractor’s estimates with the Fisher formalism to account for source covariance.

    \subsection{Photometry with Covariance Corrections}

    We refine Tractor’s uncertainty estimates by explicitly incorporating covariance among overlapping sources, following the framework of \cite{PSFportillo}. A detailed derivation is provided in Appendix~\ref{appendix:secA}.

    An instrument's optical response spreads the intrinsic flux of an observed source over an area characterized by the PSF. We denote a PSF by $p(x, y)$, where $(x,y)$ represents positions on an image. For a source $i$ with intrinsic flux $f$, the model for its observed flux is $fp_i(x, y)$ (Equation~\ref{eq:Ixy}), following \citet{PSFportillo}. Tractor maximizes the log-likelihood computed from $\chi^2$ between the model (from scaled PSFs) and the observation given by $\hat{\boldsymbol{f}} = \hat{f}(x, y)$, and outputs the maximum-likelihood flux measurement, $f^{\text{ML}}_i$, and the corresponding model $f^{\text{ML}}_ip_i(x, y)$ for a source with index $i$. We drop the sky background term in our setup, assuming that diffuse emission from structured astrophysical foregrounds or Earth's atmosphere is approximately removed through local background subtraction in early-stage image processing. Setting flux as a free parameter for forced photometry, we can compute a Gaussian likelihood, $\mathcal{L}$, by combining models for all overlapping sources and comparing with observation. Then we adopt the Fisher formalism to construct the covariance error matrix through derivatives of the log-likelihood with respect to the source fluxes. 

    For a system of two sources $u$ and $v$  it can be shown (see Appendix B of \citeauthor{PSFportillo} \citeyear{PSFportillo}) that the uncertainty is given by
        \begin{equation}
            \begin{split}
                \sigma_u^2(\theta_{\mathrm{ML}}) &= -\frac{\partial^2_v\mathrm{ln}\mathcal{L}}{\mathrm{det}\left(\mathcal{F}\right)} \\
                &= \frac{\sum_{x,y}\left[p_v(x,y)/\sigma(x,y)\right]^2}{\sum\left[p_u/\sigma\right]^2 \sum\left[p_v/\sigma\right]^2 - \sum\left[p_up_v/\sigma^2\right]^2}.
            \end{split}
        \end{equation}
        Here, $\theta_{\mathrm{ML}}$ denotes the maximum-likelihood parameters of the fit, $\mathcal{F}$ is the Fisher information matrix (FIM, Equation~\ref{eq:FIM}), and $\sigma=\sigma(x,y)$ is the per-pixel noise. The sums run over all relevant pixels $(x,y)$, defined here as those covered by the PSFs. For SPHEREx, the optical PSF is undersampled, and the pixelized model $p(x,y)$ depends on the subpixel source position, leading to variation in the images across different dithers. When two sources are sufficiently separated such that their PSF overlap is negligible, the overlapping integral vanishes in off-diagonal terms in the FIM (Equation~\ref{eq:crosspartial}). In this case, the flux uncertainty reduces to the original Tractor calculation, equivalent to the scenario of isolated sources by directly inverting diagonal terms in FIM, 
        \begin{equation}
            \sigma_{i, \mathrm{Tractor}}^2(\theta_{\mathrm{ML}}) = \left[\partial^2_i\mathrm{ln}\mathcal{L}\right]^{-1} = \left[\sum_{x,y}\left(\frac{p_i(x,y)}{\sigma(x,y)}\right)^2\right]^{-1}
            \label{eq:tractor_calc}
        \end{equation}
        
        When the PSFs overlap, the flux uncertainty will be greater than the isolated estimate, capturing the effect of statistical covariance. For the two-source scenario $\{u,v\}$ , the flux variance ratio between the blended and isolated case is 
        \begin{equation}
            \frac{\sigma_{\mathrm{blend}, u}^2}{\sigma_{\mathrm{iso}, u}^2} = \left[1 - \frac{\sum (p_u/\sigma)^2}{\sum (p_up_v/\sigma^2)} \right]^{-1} >1.
            \label{eq:fluxerr_rise}
        \end{equation}
        In this expression, the ratio is independent of the actual source fluxes, and the increased flux uncertainty is determined solely by the PSF shape and the degree of overlap. 

        In this work, we implement the generalized covariance error calculation in Tractor for an arbitrary number of overlapping sources numerically. 

    \subsection{Results of Blended Photometry}
        
\begin{figure}
        \centering
        \includegraphics[width=1.0\linewidth]{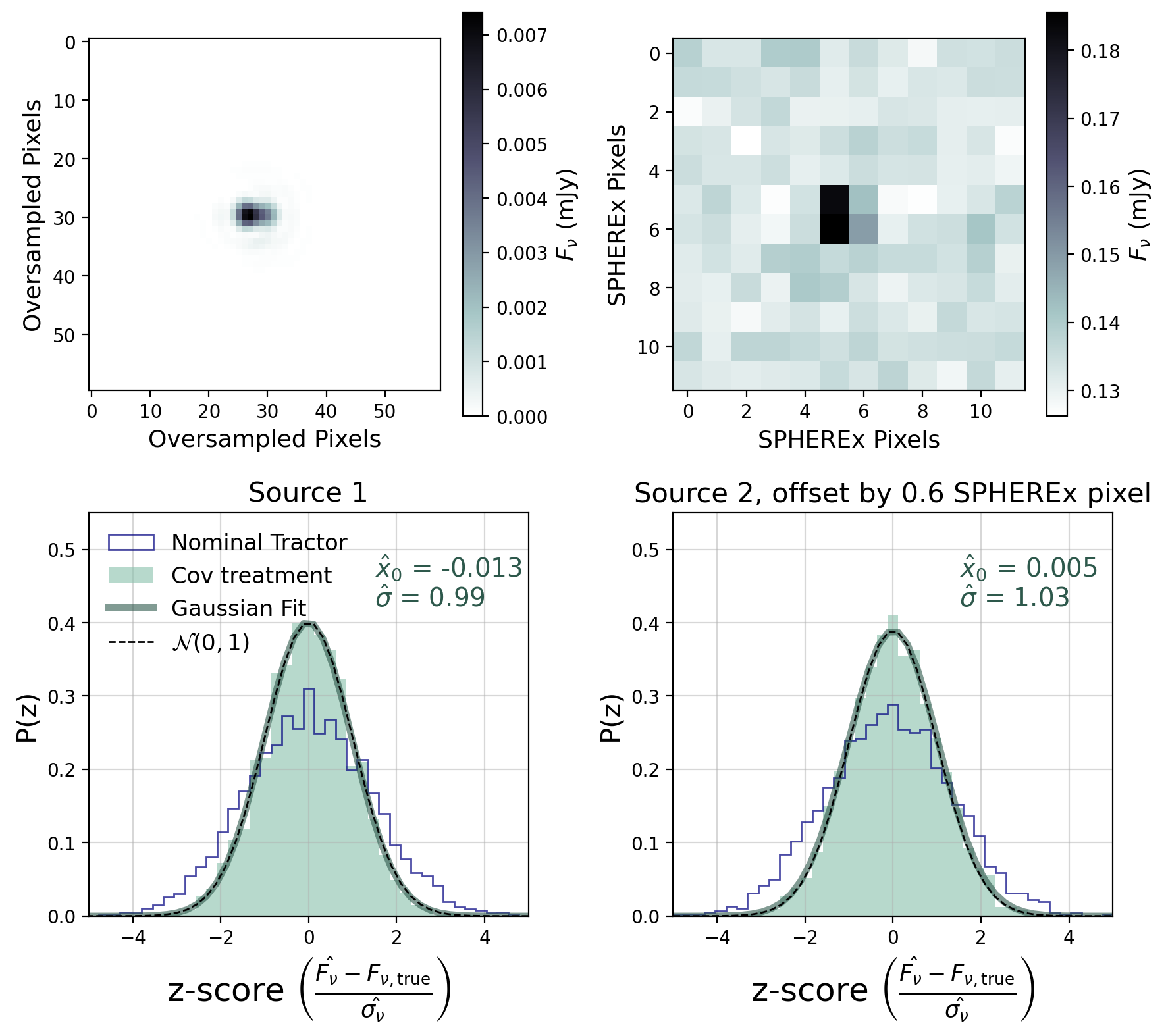}
           \caption{Tractor photometry including flux covariance. The top left panel shows a pair of blended sources in a high-resolution noiseless scene upsampled by a factor of 5 from the SPHEREx resolution $6\farcs15$. Sources are separated by 3 oversampled pixels ($3\farcs69$) and have an intrinsic flux ratio of 2:1. The image is downsampled to SPHEREx resolution with simulated noise added, as shown on the top right panel. The bottom row displays the unbiased photometric z-score based on a locally modified implementation of Tractor with covariance (light green) performed on the blended pair simultaneously, with fitted Gaussian statistics (dark green) and an ideal distribution (black dashed curve), for each source in the blended pair. This is compared with the z-score distribution from the original nominal Tractor results (blue). } 
              \label{fig:sec3/tractor_covariance}
        \end{figure}

        We validate the covariance uncertainty calculation numerically using simulated SPHEREx PSFs. For the main test shown in Figure \ref{fig:sec3/tractor_covariance}, we place two sources in a high-resolution simulated image with significant PSF overlap ($3/5$ of a SPHEREx pixel), and with flux ratio of 2:1. The image is then downsampled to SPHEREx resolution with Gaussian noise added.
        Tractor is then used to perform forced photometry by fixing the source positions and simultaneously fitting the fluxes of both sources. This process is repeated over thousands of independent noise realizations, with all other variables held constant.  
        
        Through this Monte Carlo procedure, we can estimate the flux z-score distribution, assuming both uncorrelated and correlated uncertainties. In Figure \ref{fig:sec3/tractor_covariance} we show the results of this comparison. Despite significant blending, we observe no bias when incorporating the source-source flux covariance: the z-score distributions are consistent with $\mathcal{N}(0,1)$, zero mean and unit width. In contrast, using isolated source uncertainties leads to an underestimation of errors and wider z-score distributions. We have also tested a wide range of flux ratios (1:100 to 100:1), separations (down to one-tenth of a SPHEREx pixel), and subpixel positions with varying PSF orientations; in all cases, the recovered fluxes are unbiased. Furthermore, we extended this test to crowded scenes containing up to $\sim\!30$ sources randomly distributed within a $\sim\!1\ \mathrm{arcmin}^2$ image, simultaneously fitted using Tractor. Flux estimates remain statistically unbiased, even in cases of significant blending. 

        Our simulation assumes perfect astrometry, which is not realistic due to positional errors in both external catalogs and internal calibration. In the forced photometry of isolated sources, this leads to a negative flux bias \citep{PSFportillo}. Among blended sources, however, astrometry errors do not necessarily bias all sources negatively, as the local $\chi^2$ minimization involves combined flux contributions. The exact impact also depends on the orientation and magnitude of the astrometric offsets in blended scenes.

    \subsection{Photometric Error Inflation with Target Density}

    \subsubsection{Flux Error Dependence on Source Proximity}
    \label{sec3/zodi}
    Source blending increases estimated flux uncertainties, which degrades redshift performance compared to isolated sources. In SPHEREx, the dominant source of noise is photon noise from the ZL. For the blended case, at the pixel level $(x, y)$, photon noise is 
    \begin{equation}
        \sigma_{\mathrm{phot, blend}}(x, y) \propto \sqrt{\sum_i^{N_\mathrm{src}} p_i(x,y)f_i +C},
        \label{eq:scale_fraction1}
    \end{equation} 
    where $p_i(x,y)f_i$ is the fractional flux from the PSF of the $i$-th source. The constant $C$ represents the ZL photon noise, which in principle depends on position and time $(x,y,t)$, but can be treated as a constant on the angular scales of individual sources. For an isolated source $i$, this reduces to, 
    \begin{equation}
         \sigma_{\mathrm{phot, isol}}(x, y) \propto \sqrt{p(x,y)f + C}.
        \label{eq:scale_fraction2}
    \end{equation}
    The ratio $\frac{\sigma_{\mathrm{phot, blend}}}{\sigma_{\mathrm{phot, iso}}}$ thus depends not only on PSF overlap and source separation, but also on the relative fluxes of the contributing sources. However, the background term $C$, dominated by the ZL, exceeds the photon noise contribution from faint sources. To quantify this, we compute the minimum source flux at which source photon noise becomes significant, defined as the point where forced-photometry uncertainty (including pixelization effects) exceeds the instrument noise by 10\%: $\delta f/\delta f_{\rm ZL} = 1.1$. For SPHEREx, this threshold corresponds to magnitudes of 18.8 at $0.75\ \mu\rm m$ and 17.4 at $5.0\ \mu \rm m$, based on the sky-averaged ZL estimates. This is consistent with stronger ZL at longer wavelengths. Since most galaxies in our sample are fainter than these limits, their photometric uncertainties are largely set by the relatively uniform ZL background, with minimal dependence on source flux. 

\begin{figure}
        \centering
        \includegraphics[width=0.95\linewidth]{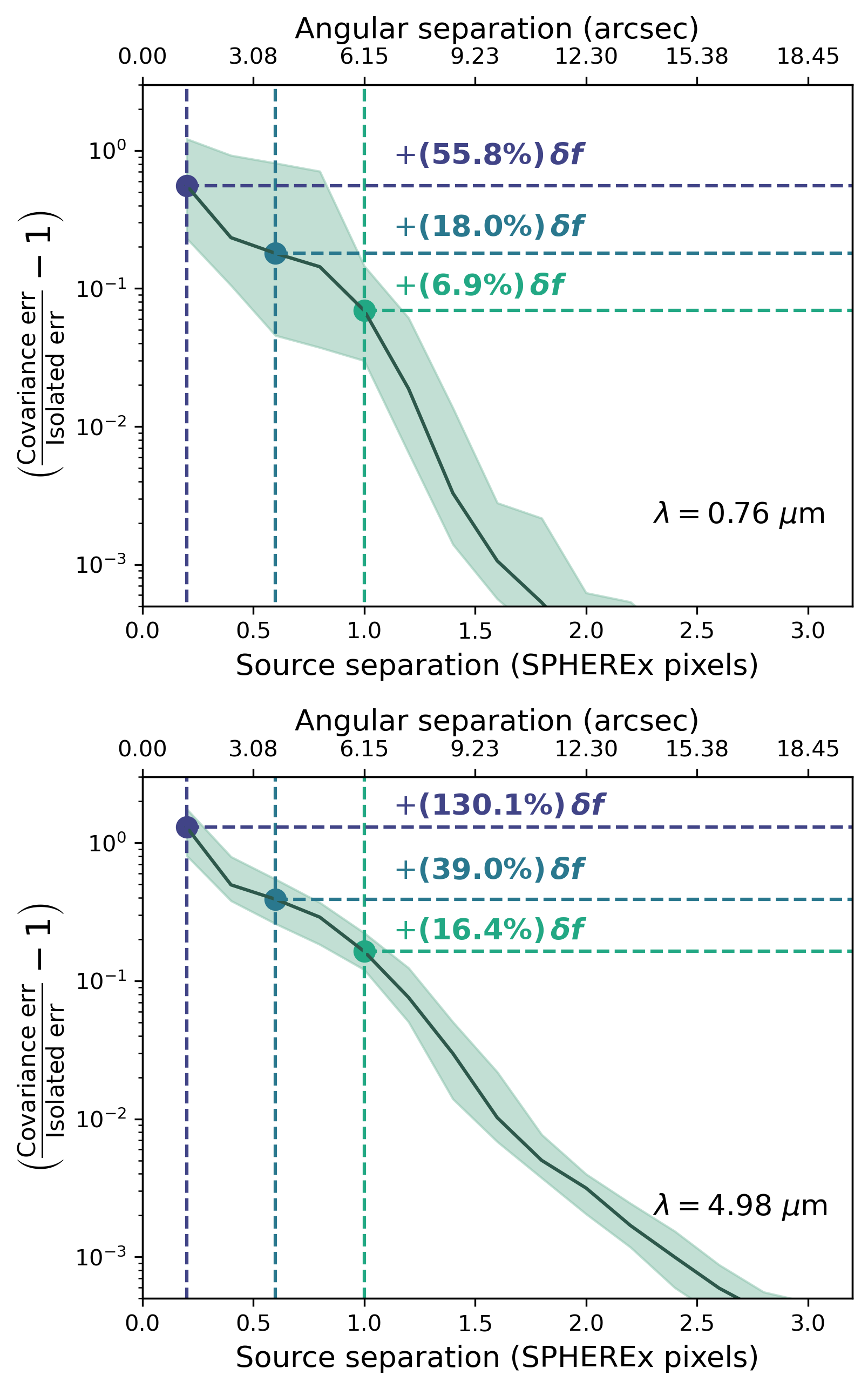}
           \caption{Fractional excess in flux uncertainty for a pair of blended sources (flux ratio 2:1) as a function of their separation. Photon noise and readout noise are included, with subpixel sampling achieved by averaging over PSF orientations. Both sources are below the ZL photon noise floor, so the choice of which source to show is unimportant; here we show the fainter one. At shorter wavelengths (\textbf{top}), the PSF is more asymmetric, resulting in a broader 16–84\% variation band (shaded light green) around the median curve. At longer wavelengths (\textbf{bottom}), the median flux inflation (black) is higher due to larger diffraction and thus greater PSF overlap. The fractional increase in flux uncertainty, $\delta f$, for separations of 1.0, 0.6, and 0.2 SPHEREx pixels is labeled.}
              \label{fig:sec3/flux_err_increase}
    \end{figure}

    To illustrate the impact of blending on flux uncertainties, we simulate two artificial sources with realistic noise. Figure~\ref{fig:sec3/flux_err_increase} shows how this uncertainty grows as the source separation decreases, in two SPHEREx spectral channels. At longer wavelengths, increased diffraction leads to a larger PSF and larger uncertainty, while at shorter wavelengths, PSF asymmetry causes greater scatter depending on the source position. Therefore, blending between photometered sources results in a non-uniform increase in flux uncertainties while remaining unbiased if no astrometry errors are present. 

    \subsubsection{Flux Uncertainty vs. Selection Depth}
    \label{sec:3.4.2}
    We have shown that flux uncertainty increases for individual targets as proximity to neighboring sources decreases. Since this is directly related to the number density of selected targets, we now quantify how ensemble-level flux uncertainties scale with selection depth, i.e., target density. This study motivates careful investigation of the magnitude threshold used to define the photometric sample. 

    We apply five LS $z$-band magnitude cuts to COSMOS galaxies in extragalactic fields only (LS-$z<21.3,\ 22.0,\ 22.7,\ 23.4$), chosen as a depth threshold to mimic the cosmology selection in Equation~\ref{eq:cut}. These cuts yield target densities of $n=2.6,\ 5.1,\ 9.8,\ 17.1\ \mathrm{arcmin}^{-2}$, covering the density in the current cosmology sample. For each cut, we perform:
    \begin{itemize}
        \item Blended photometry (denoted $B$) using Tractor with full covariance treatments in realistic scenes.
        
        \item Isolated photometry (denoted $\mathrm{iso}$), photometering each target individually and in isolation without covariance.
    \end{itemize}
    
\begin{figure}
        \centering
        \includegraphics[width=1.0\linewidth]{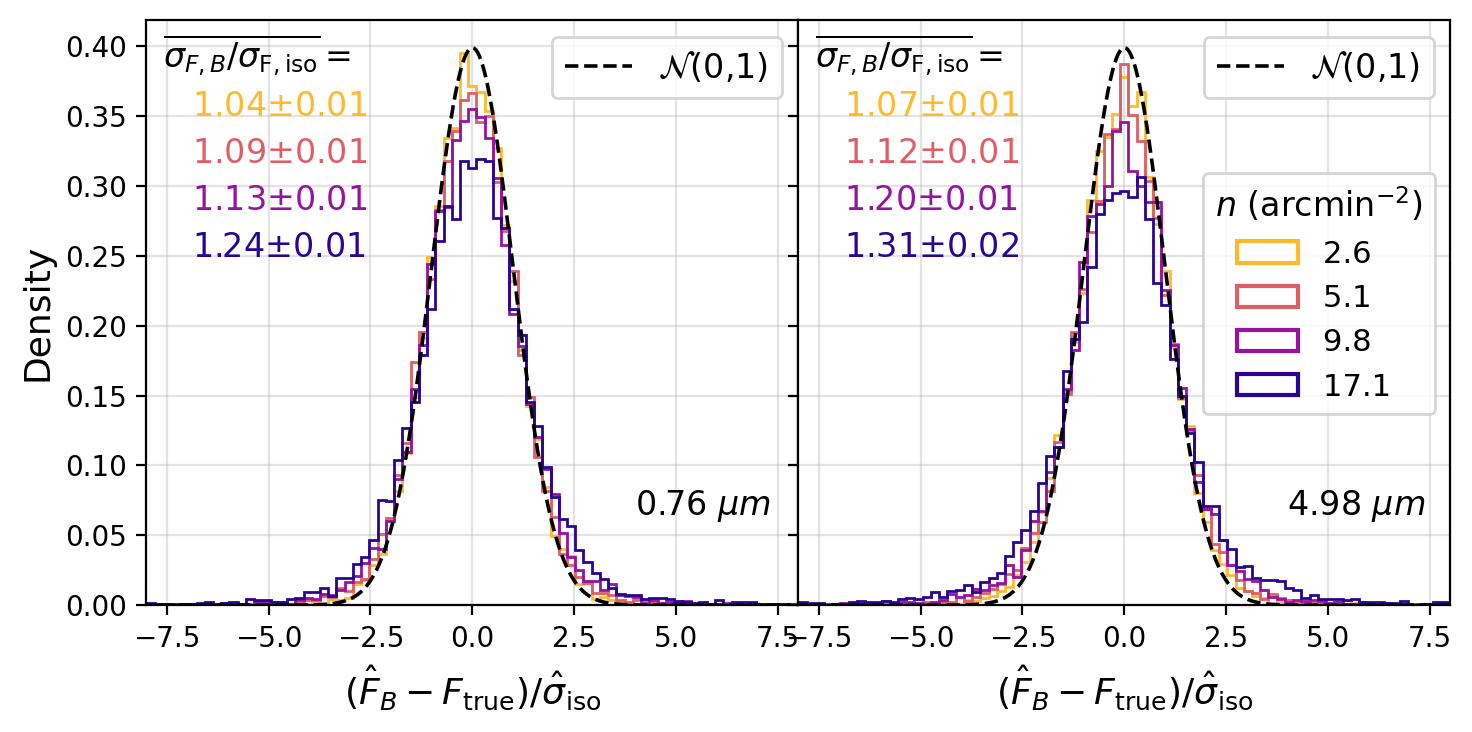}
        \caption{Ensemble flux error inflation among targeted galaxies. For each $z$-magnitude cut (LS-$z<21.3,\ 22.0,\ 22.7,\ 23.4$) that corresponds to a target number density (color-coded), we compare the true photometric error from the covariance mode $\hat{F}_B$ to the naive isolated flux error $\hat{\sigma}_{\mathrm{iso}}$. Distributions of $(\hat{F}_B - F_{\text{true}})/\hat{\sigma}_{\mathrm{iso}}$ are shown, with a standard normal (black) for reference. The broadening is quantified by the NMAD, labeled as $\overline{\sigma_{F,B} / \sigma_{\mathrm{F, iso}}}$ for each sample of galaxies. Results are shown for two SPHEREx channels: 0.76 (left) and $4.98\ \mu \mathrm{m}$ (right). }
        \label{fig:sec3/xzscore_B}
    \end{figure}

    The ensemble flux error increases as a direct result of increased blending. 
    Figure~\ref{fig:sec3/xzscore_B} shows the distribution of the ratio of blended flux error to isolated error, which broadens with increasing target density. We characterize the flux error inflation using the normalized median absolute deviation (NMAD) of this ratio: $\overline{\sigma_{F,B} / \sigma_{\mathrm{F, iso}}}$. We compute bootstrap uncertainties by resampling the galaxy sample with random replacement and taking the standard deviation of the resulting NMAD values.

\begin{figure}
        \centering
        \includegraphics[width=0.9\linewidth]{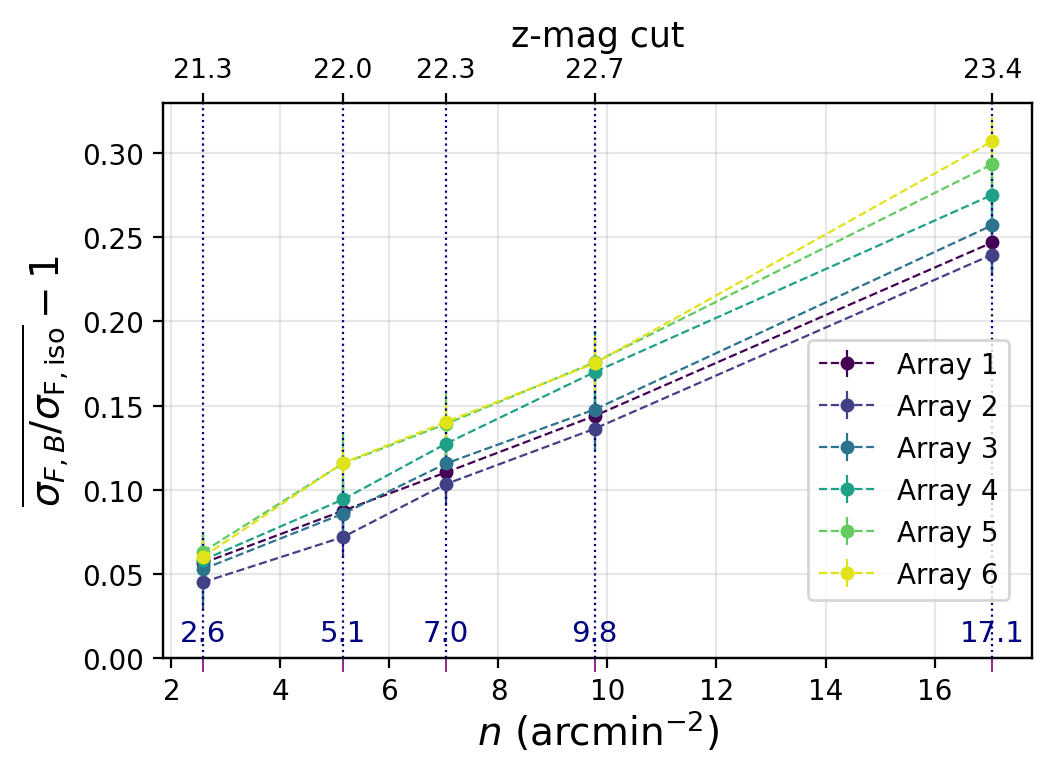}
        \caption{Fractional increase in flux error due to blending vs. target number density. Each set of points corresponds to a $z$-magnitude cut, with associated number densities on the lower $x$-axis (navy) and $z$-magnitude on the upper axis. The LS-$z < 22.3$ cut matches the number density of our fiducial cosmology sample ($\sim\!7.0$ arcmin$^{-2}$).}
        \label{fig:sec3/sigmaF_B}
    \end{figure}

    The distributions in Figure~\ref{fig:sec3/xzscore_B} illustrate how we derive the summary statistics shown in Figure~\ref{fig:sec3/sigmaF_B}. We present the characteristic inflation of flux errors of selected samples versus target density. For the current cosmology color-magnitude selection (Equation~\ref{eq:cut}), we observe a $\sim\!11\%$ ($0.75-1.10\ \mu\mathrm{m}$) to $\sim\!14\%$ ($4.40 - 5.00\ \mu\mathrm{m}$) increase in the flux uncertainty compared with the isolated case. This increase remains relatively consistent between full-sky and deep-field survey depths, as illustrated in Figures~\ref{fig:sec5/BS/sigz_acc_full} and \ref{fig:sec5/BS/sigz_acc_deep}, primarily because the ZL dominates the noise budget (see Section~\ref{sec3/zodi}). A larger degradation at longer wavelengths reflects the broader PSFs. 
    
\section{Modeling Spectral Confusion}
\label{sec:lib_method}
We construct a ``spectral confusion library" from untargeted galaxies with simulated SPHEREx observations to model and characterize spectral confusion noise. We analyze how this noise scales with both survey sensitivity and target selection depth, and assess enhancements due to clustering. These tools allow photometric-level injection of confusion into simulated data for redshift studies.

    \subsection{Construction of Spectral Confusion Library}

    We model spectral confusion by constructing a \emph{spectral confusion library} from untargeted galaxies fainter than our fiducial selection (Equation~\ref{eq:cut}), down to $m = 25$ in the $i$-band, which is the limit of the COSMOS2020 catalog we use, approximately three magnitudes fainter than the forecast SPHEREx deep-field $5\sigma$ point source sensitivity. The depth of the COSMOS catalog ensures that we capture the cumulative contribution of faint galaxies to confusion. In trial tests, we find that integrated contamination from these untargeted galaxies converges by magnitude LS-$z\sim24.5$, indicating the catalog is sufficiently deep.

    \begin{figure*}[ht]
        \centering
        \begin{tikzpicture}[
            node distance=2cm and 7cm, 
            every node/.style={minimum width=3cm, minimum height=1.5cm}, 
            every path/.style={thick} 
        ]

        \node[xshift=-6cm, yshift=2.6cm] (mid1) {\includegraphics[width=8.5cm]{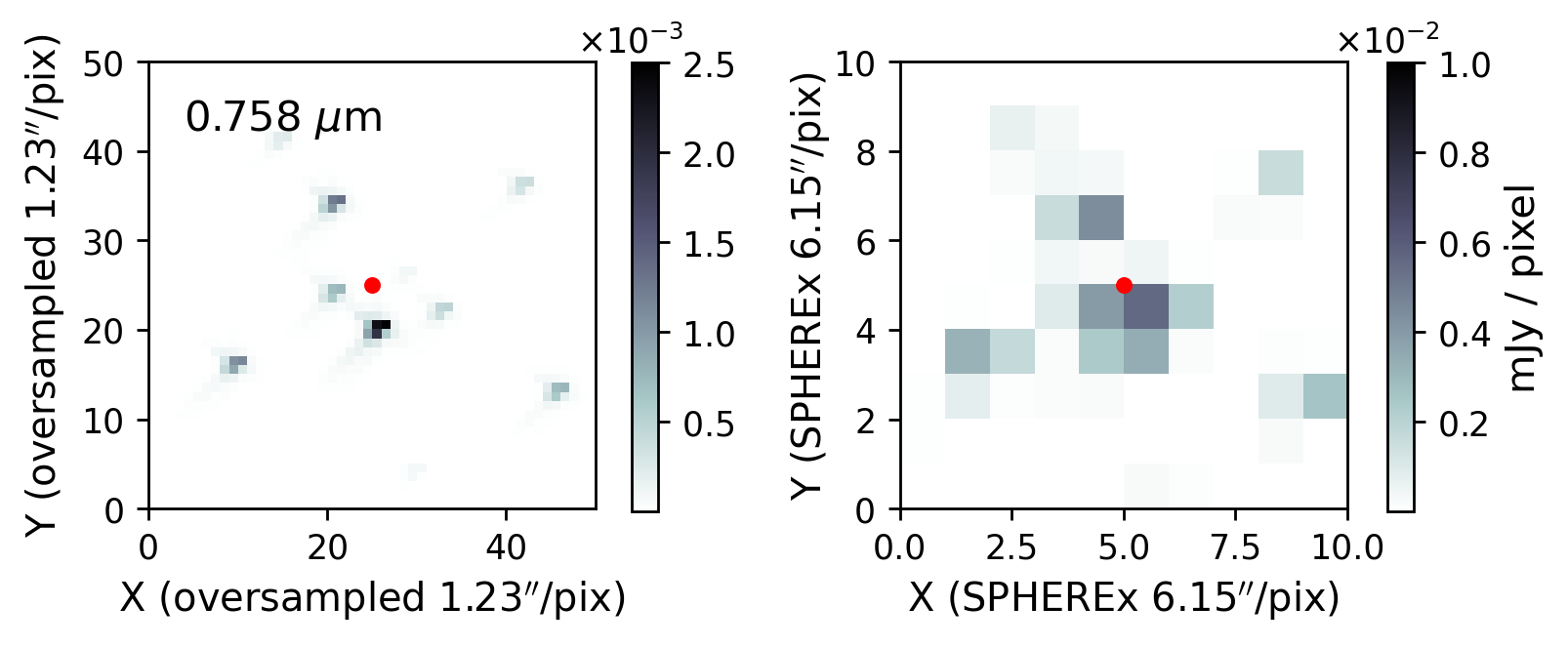}};

        \node[xshift=-6cm, yshift=0cm, draw=none] (mid2) {
            \begin{tabular}{c}
                \rotatebox{90}{\textbf{...}} \\
                Other spectral channels \\
                \rotatebox{90}{\textbf{...}}
            \end{tabular}
        };

        \node[xshift=-6cm, yshift=-2.6cm] (mid3) {\includegraphics[width=8.5cm]{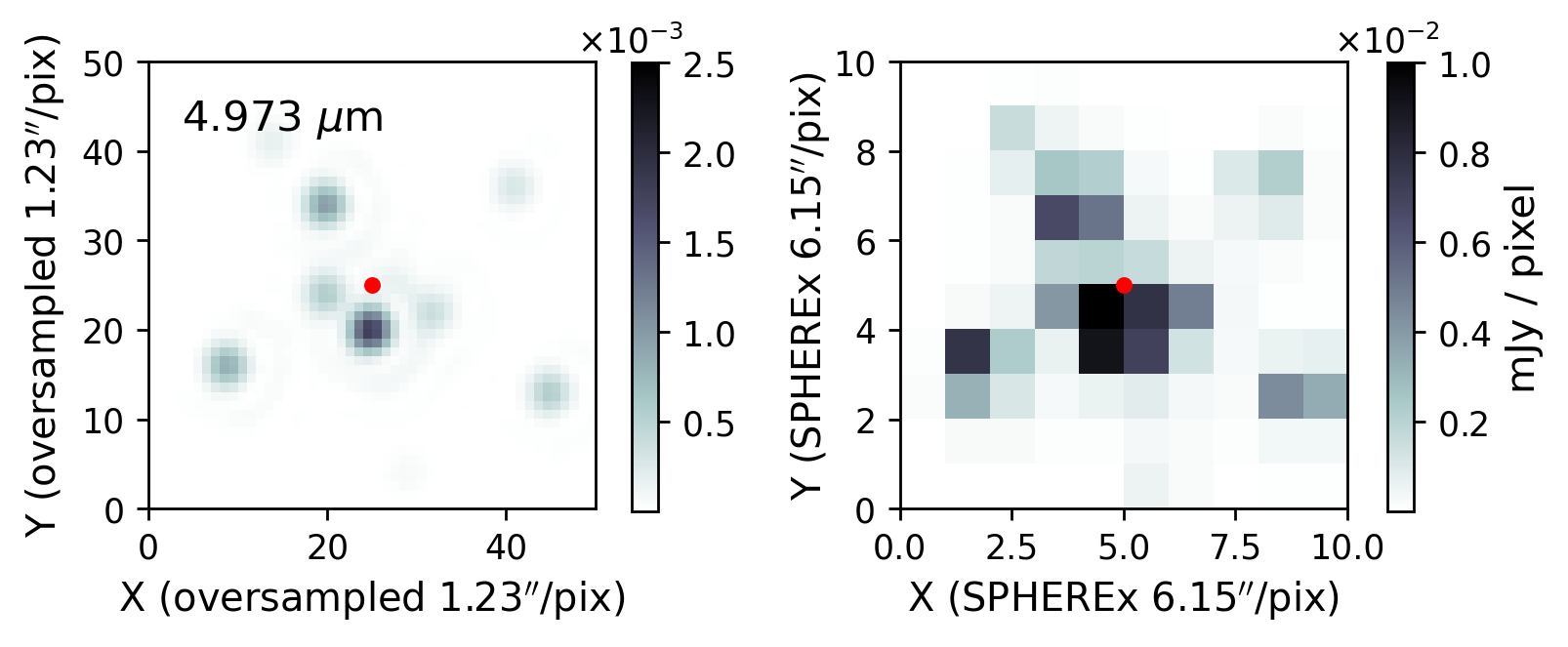}};
        
        \node[draw][right=of mid2, xshift=-2cm] (right) {\includegraphics[width=3.5cm]{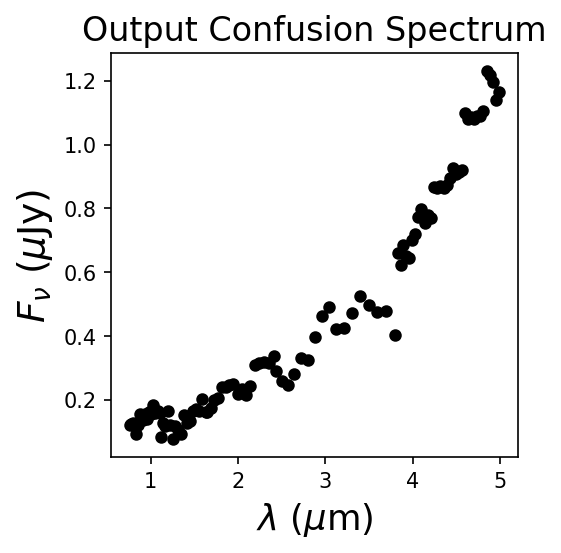}};

        \draw[decorate,decoration={brace,amplitude=6pt,raise=1pt}, thick] 
            ([yshift=2cm]mid1.south east) -- ([yshift=-2cm]mid3.north east)
            node[left,xshift=0.7cm,midway]{};  
        
        \draw[->] ([xshift=2.5cm]mid2.east) -- ([xshift=-0.3cm]right.west) node[midway, above] {};
        
        \end{tikzpicture}
        \caption{Flowchart of confusion library construction: A random position is selected within the COSMOS field and force-photometered (fixed at the selected position) alongside nearby untargeted sources, which are rendered into simulated images based on the SPHEREx survey plan. The red point indicates the force-photometered position and the underlying grayscale pixel map shows only untargeted (excluded by the selection in Equation~\ref{eq:cut}) sources from the COSMOS catalog.  The left column shows oversampled images at 5$\times$ the native SPHEREx resolution, while the middle column shows native-resolution ($6\farcs15$) images for different spectral channels. The right column presents the recovered flux contribution from the untargeted sources in the absence of other noise sources as an example. Thousands of such realizations comprise the confusion library.}
        \label{fig:sec4/confusion_lib_construct}
    \end{figure*}
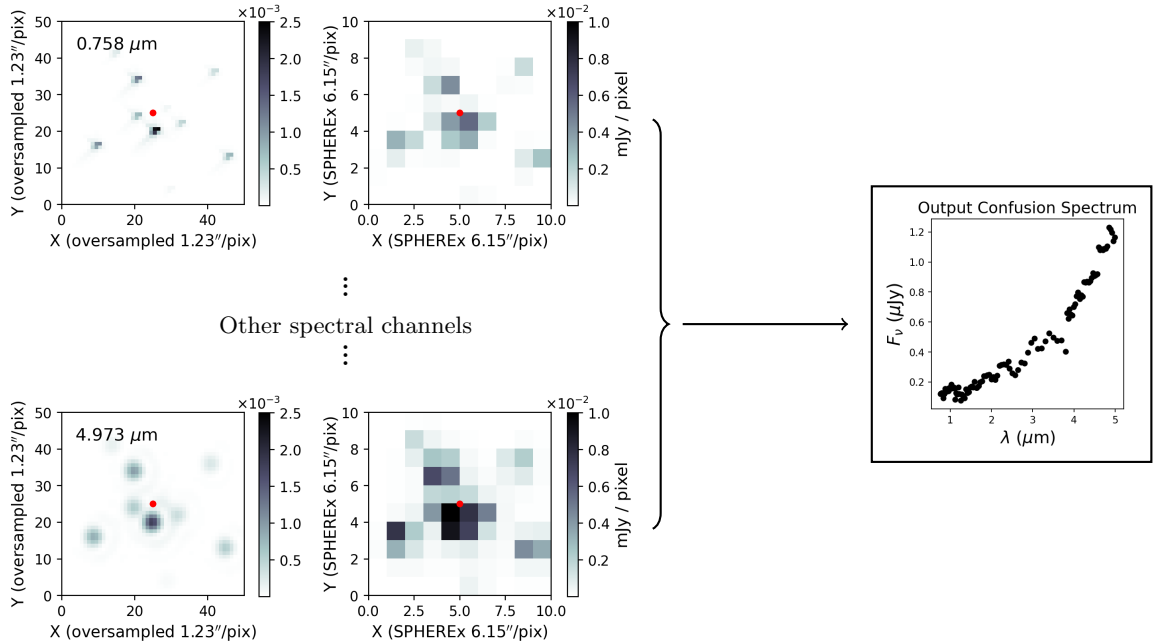
     
    The workflow for constructing the confusion library is illustrated in Fig.~\ref{fig:sec4/confusion_lib_construct}. The idea is to isolate flux contributions from faint, untargeted sources and compile a comprehensive library of 6000 ``confusion spectra". We obtain each confusion spectrum by randomly selecting coordinates in the COSMOS field, then simulating the combined flux from nearby faint, untargeted sources by adding them onto the cutout using their high-resolution SEDs. Due to the random nature of this library, we refer to it as the \textit{stochastic library}. Figure~\ref{fig:sec4/confusion_lib_construct} illustrates the construction of this library at the beginning and end of the SPHEREx wavelength range. 
    
    To isolate the impact of source confusion, all other noise components — such as the ZL, read noise, dark current, and photon noise — are effectively disabled by setting the per-pixel noise variance to a negligible floor (specifically, $10^{-15}$ times the nominal noise). This ensures numerical stability in Tractor while suppressing any influence from other noise sources. Tractor photometry is then performed by fixing the model at the selected coordinates (red points in Figure~\ref{fig:sec4/confusion_lib_construct}) and fitting the central blank spot to extract confusion-induced fluxes. This process is applied across all SPHEREx observations following the 2-year survey plan.

    Within the COSMOS field, we find that the confusion variance converges after roughly $3000–4000$ realization. We therefore adopt 6000 realizations to ensure robust convergence. By removing other noise components, we obtain unbiased confusion estimates, independent of the number of observations.

    \subsection{Variation Across the Confusion Library}

\begin{figure}
        \centering
        \includegraphics[width=0.9\linewidth]{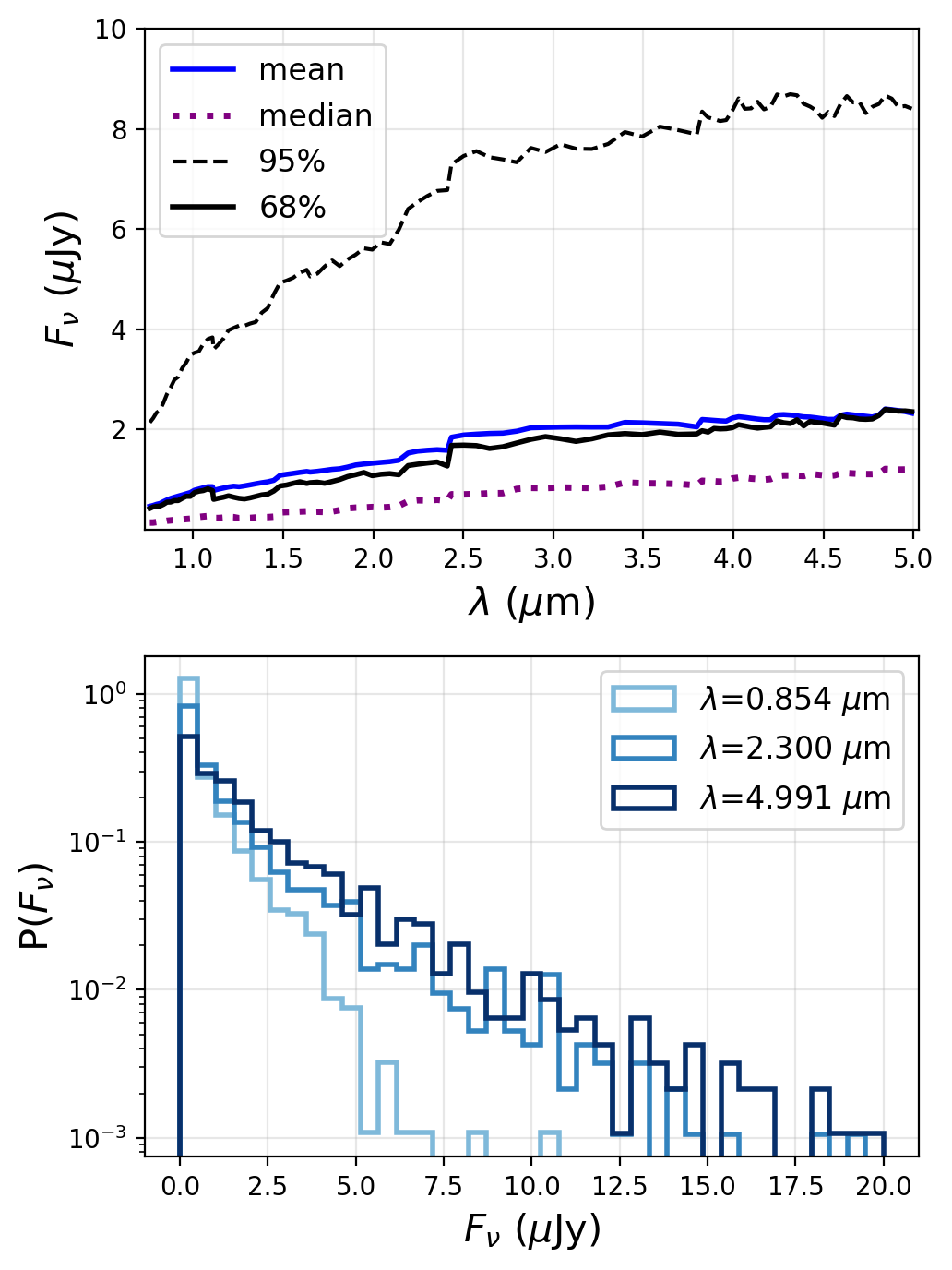}
           \caption{Distribution of the confusion library. \textbf{Top Panel:} the mean (blue), median (purple), and quantiles at 68\% and 95\% (black) measured from zero to capture outliers, are overlaid for each spectral channel. \textbf{Bottom Panel:} Flux density distribution of confusion library in three selected channels.}
              \label{fig:sec4/confusion_lib}
    \end{figure}

    After library construction, we characterize the distribution of confusion fluxes across spectral channels in Figure~\ref{fig:sec4/confusion_lib}.
    A heavy-tail distribution is observed across all spectral channels. In the three selected channels, the distribution becomes wider at longer wavelengths, consistent with higher diffraction, broader PSFs, and thus more extended spectral confusion. The wavelength-dependent spectral shape introduces a color bias, which is partially driven by our cosmology selection. 

\begin{figure}
        \centering
        \includegraphics[width=1.0\linewidth]{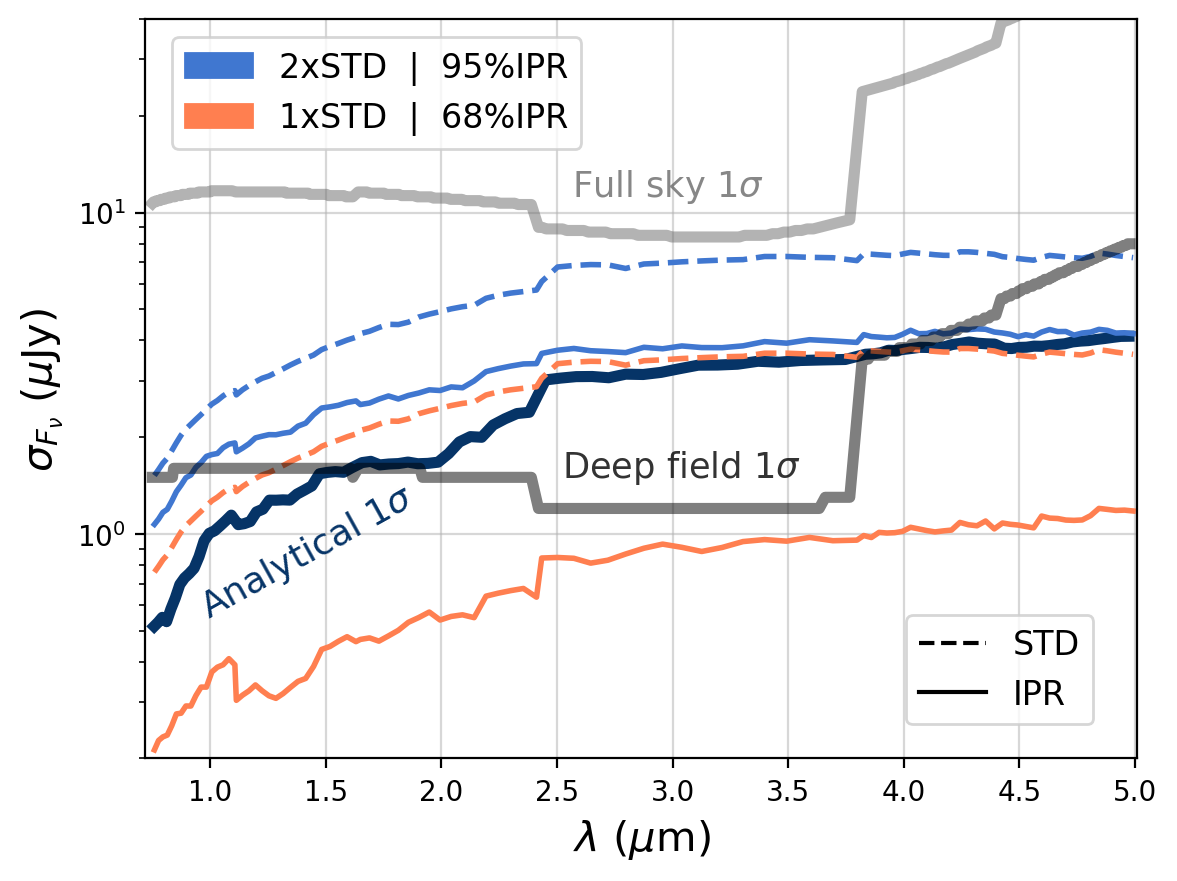}
           \caption{Confusion library variation compared with SPHEREx 1-$\sigma$ sensitivity in the deep field and all sky (grey): $1\sigma$ (orange) and $2\sigma$ (blue) variation in each channel is displayed with two metrics, STD (dashed) and IPR (solid lines). STD overall reports larger variation due to its susceptibility to brighter tails, whereas IPR better captures the ensemble behavior near the peak. The analytical $\sigma_{\mathrm{conf}}$ from Equation~\ref{eq:analyt_confusion} (navy) roughly coincides with the STD $1\sigma$. The wavelength-dependent, correlated features of this variation introduce a color bias in the spectro-photometry.
                   }
              \label{fig:sec4/lib_variation}
    \end{figure}

    We characterize the variation using the standard deviation (STD) and the interpercentile ranges (IPRs), e.g., the $16-84\%$ and $2.5-97.5\%$ ranges, corresponding approximately to 1$\sigma$ and 2$\sigma$ intervals of the flux distribution. Compared to the standard deviation, the IPR more accurately captures the shape of heavy-tailed, asymmetric distributions. Figure~\ref{fig:sec4/lib_variation} compares the 68\% and 95\% STD and IPR intervals with SPHEREx sensitivity in the deep field and full sky. While STD captures overall spread, it can be inflated by outliers in skewed or heavy-tailed distributions. In contrast, the 68\% IPR is less sensitive to such tails.
    Figure~\ref{fig:sec4/lib_variation} shows that the 68\% IPR confusion variation falls below the SPHEREx 1-$\sigma$ sensitivity in the deep field and well below the full-sky sensitivity. However, the 95\% IPR and STD both exceed the deep-field sensitivity. The variation introduces a spectral (color) bias into the photometry, which can systematically bias redshift estimates. 

    \subsection{ Validation of the Confusion Library}
    \label{subsec:lib_validation}

    Before applying the spectral confusion library to redshift measurements, we assess the spectral variation by comparing it against a ``clustering library" and a traditional analytical approach. We construct the clustering library primarily for comparison, to validate that our stochastic confusion library captures the first-order effects of enhanced confusion due to spatial clustering.

    Because the confusion library is built using randomly selected sky positions, it may underestimate enhanced confusion from faint satellite galaxies clustered around brighter primaries, or from chance line-of-sight projections. Such clustering, manifested in galaxy power spectra \citep{scherrer98, barcons92, tegmark04}, can lead to elevated confusion levels beyond those captured by our baseline assumption of randomness.

    The workflow for constructing the \emph{clustering} library follows that of Figure~\ref{fig:sec4/confusion_lib_construct}, with one key difference: 
    instead of randomly sampling sky coordinates, we perform photometry at the actual positions of real targeted sources, denoted $(\alpha_0, \delta_0)$, where $\alpha_0$ and $\delta_0$ are the RA and DEC of each source that passes the selection in Eq.~\ref{eq:cut}.
    We exclude the flux of the primary source itself at each location to isolate confusion arising purely from neighboring untargeted galaxies. This approach preserves spatial clustering and allows us to quantify the enhanced confusion contribution from physically correlated structures.

    The two libraries are consistent to within about 5\% on average. The small excess in the clustering case likely reflects additional contributions from small-scale galaxy clustering. Given the level of agreement, the stochastic library provides a reliable representation of confusion for our purposes.
    
    We note that our confusion library does not explicitly include diffuse background components such as the ZL and diffuse Galactic light (DGL). In our analysis these are treated as smooth backgrounds that can be subtracted in point source photometry. For ZL, previous work indicates that fluctuations on sub-arcminute scales are negligible \citet{pyo12}. However, spatial variations below a few arcminutes may introduce additional confusion \citet{jeong05}, though such residuals are expected to have only a minor impact in deep fields.

    To complement these numerical tests, we compute the analytical confusion noise using the common definition in \cite{condon, bethermin24, helou&beichman90},
    \begin{equation}
        \sigma_{\mathrm{conf}}^2 = \iint b^2 \mathrm{d}\Omega \int_0^{F_\mathrm{untarg}} F_{\nu}^2 \frac{\rm{d}N}{\rm{d}F_{\nu}} \mathrm{d}F_{\nu},
        \label{eq:analyt_confusion}
    \end{equation}
    where $b$ is the beam normalized to unity at its peak, and $\frac{\rm{d}N}{\rm{d}F_{\nu}}$ is the galaxy number density per flux bin per steradian. Note that Equation~\ref{eq:analyt_confusion} does not account for clustering effects, but has been widely used in source detection. Using high-resolution galaxy SEDs and widths of pixelized SPHEREx PSFs, we evaluate $\sigma_{\mathrm{conf}}$ for each SPHEREx channel, and compare the result with numerical estimates. We integrate over the fluxes of all galaxies in the untargeted catalog fainter than the selection cut, $F_{\rm untarg}$. This analytical estimate is directly compared with our numerical simulations, which also use the COSMOS2020 catalog down to the same faint limit, ensuring consistency between the two approaches.

    The analytical confusion estimate shown in Figure~\ref{fig:sec4/lib_variation} agrees well with the numerical simulations in terms of the STD metric, as expected from the second-moment formalism in Equation~\ref{eq:analyt_confusion}, which sums up Poissonian contributions.

    \subsection{Spectral Confusion Dependence on Target Selection and Survey Depth}
    \label{sec4.4}
    
    \begin{figure}
        \centering
        \includegraphics[width=1.0\linewidth]{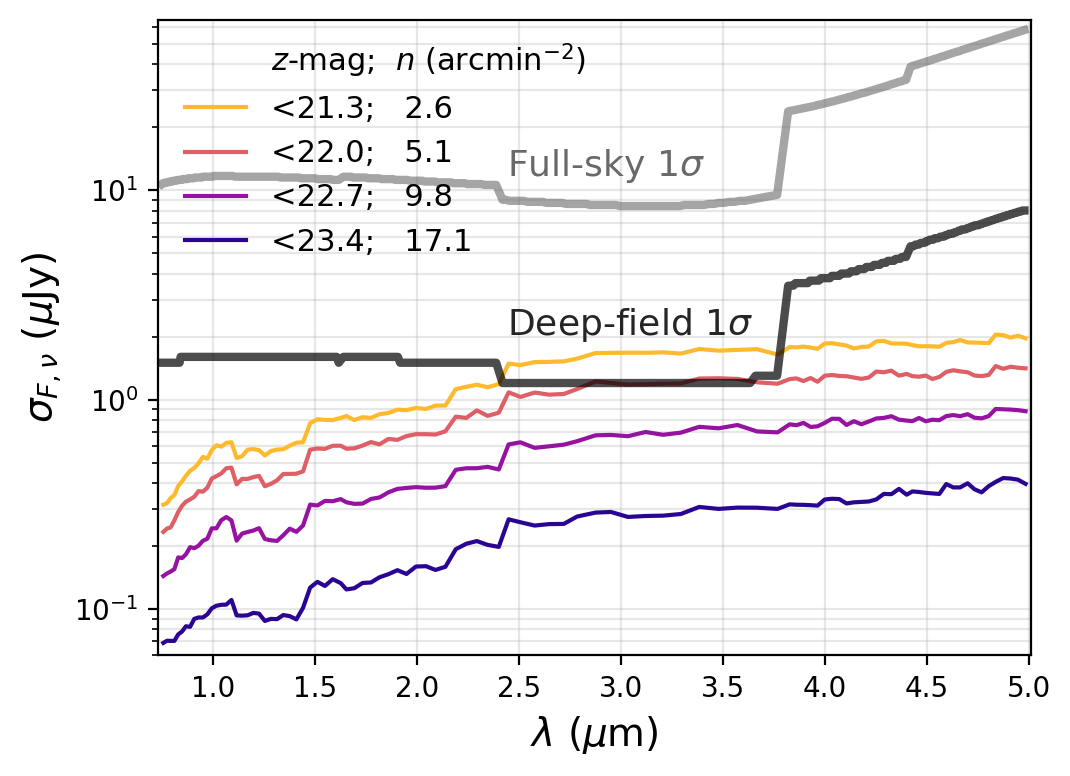}
        \caption{Confusion noise versus selection depth. The 68\% IPR of each confusion library, corresponding to different LS-$z$ magnitude cuts, is shown as a function of wavelength. SPHEREx full-sky and deep-field $1\sigma$ point source sensitivities are over-plotted for comparison. }
        \label{fig:sec4/sigmaS_specconf}
    \end{figure}

    We examine how spectral confusion depends on both selection depth and survey sensitivity. For each LS-$z$ magnitude cut defined in Section~\ref{sec:3.4.2} where we keep the LS-$z-\rm W1$ color criteria, we generate a spectral confusion library from untargeted background galaxies. Figure~\ref{fig:sec4/sigmaS_specconf} shows the 68\% IPR of these confusion libraries, compared with SPHEREx full-sky and deep-field sensitivities. The total uncertainty depends jointly on the target selection depth and survey sensitivity,  $\sigma_{\mathrm{tot}}^2 = \sigma_{\mathrm{inst}}^2 + \sigma_{\mathrm{conf}}^2$. In deep fields, where $\sigma_{\mathrm{inst}}$ is lower, confusion noise becomes a more significant contributor to the total uncertainty. 

\begin{figure}
        \centering
        \includegraphics[width=0.9\linewidth]{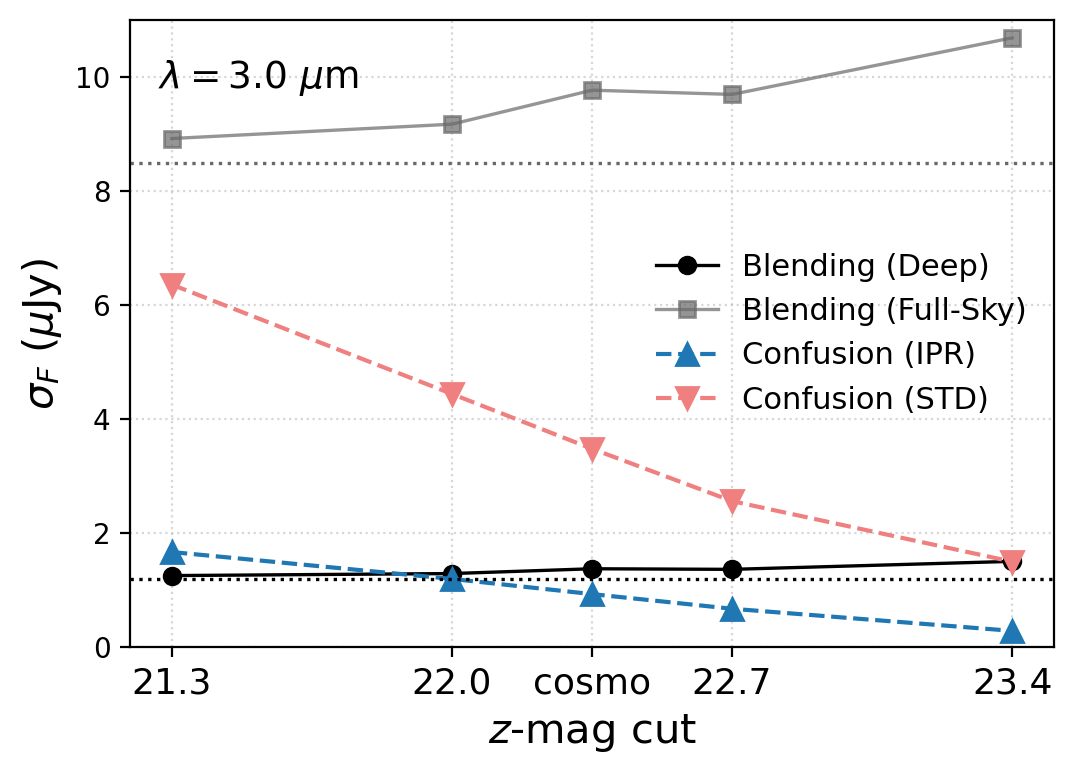}
        \caption{Comparison of photometric uncertainty due to blending and spectral confusion at $3\ \mu\rm m$ as a function of $z$-band magnitude cuts. The isolated full-sky and deep-field sensitivities (dotted horizontal lines) are scaled by the blending-induced flux inflation from Figure~\ref{fig:sec3/sigmaF_B}. Spectral confusion is shown using the 68\% IPR and STD as triangles. The cosmology sample selection is labeled on the x-axis as `cosmo'. }
        \label{fig:sec4/compare_BS_fluxerr}
    \end{figure}

    We also compare the contributions of spectral confusion and blending to the flux uncertainties in Figure~\ref{fig:sec4/compare_BS_fluxerr}. In this comparison, blending is modeled as a multiplicative increase in instrument sensitivity, based on Figure~\ref{fig:sec3/sigmaF_B}, while spectral confusion is treated as an additive contribution. At full-sky depth, blending dominates and grows with deeper cuts, whereas in the deep field, blending and confusion contribute comparably. Blending rises with deeper cuts due to more targeted sources, while spectral confusion decreases as fewer bright untargeted sources remain. This illustrates a trade-off that depends on target selection.

    To quantify the impact of selection depth and survey depth on spectrophotometric redshift measurements, we inject these confusion spectra for each $z$-magnitude cut into simulated SPHEREx photometry, applying both full-sky and deep-field noise levels. 
    


    \subsection{Confusion Injection}
    \label{sec:injection}
    
    We inject randomly drawn confusion spectra into galaxy photometry by modifying both flux and flux uncertainties as follows, 
    \begin{equation}
        F_{\nu, \mathrm{tot}}(\lambda) = F_{\nu, \mathrm{og}}(\lambda) + F_{\nu, \mathrm{C}_i}(\lambda) - F_{\nu,\mathrm{C}_{avg}}(\lambda),
        \label{eq:injection_F}
    \end{equation}
    \begin{equation}
        \sigma_{\nu, \mathrm{tot}}(\lambda) = \sqrt{\sigma_{\nu, \mathrm{og}}(\lambda)^2 + \sigma_{\nu, \mathrm{C}_{1\sigma}}(\lambda)^2}.
        \label{eq:injection_sigma}
    \end{equation}
    A confusion-perturbed spectrum $F_{\nu, \mathrm{tot}}(\lambda)$ is the sum of the original one $F_{\nu, \mathrm{og}}(\lambda)$ and a randomly sampled confusion realization $F_{\nu, \mathrm{C}_i}(\lambda)$. The average confusion in the library $F_{\nu,\mathrm{C}_{avg}}(\lambda)$ is subtracted to mimic background removal in SPHEREx Level-3 processing. Accordingly, flux uncertainties are inflated by the confusion library's 1-$\sigma$ variation, $\sigma_{\nu, \mathrm{C}_{1\sigma}}(\lambda)$, added in quadrature to the original errors $\sigma_{\nu, \mathrm{og}}(\lambda)$. The relative impact of confusion depends on instrument sensitivity: it is negligible in the full sky where instrument noise dominates but becomes more significant in the deep field with lower noise levels.

\section{Results: Impact on Redshift}
\label{sec5}

Using the photometric tools developed for blending and spectral confusion, we now assess their impact on redshift performance.

\subsection{Blending}
\label{sec5:res_B}

    \begin{figure}
        \centering
        \includegraphics[width=1.0\linewidth]{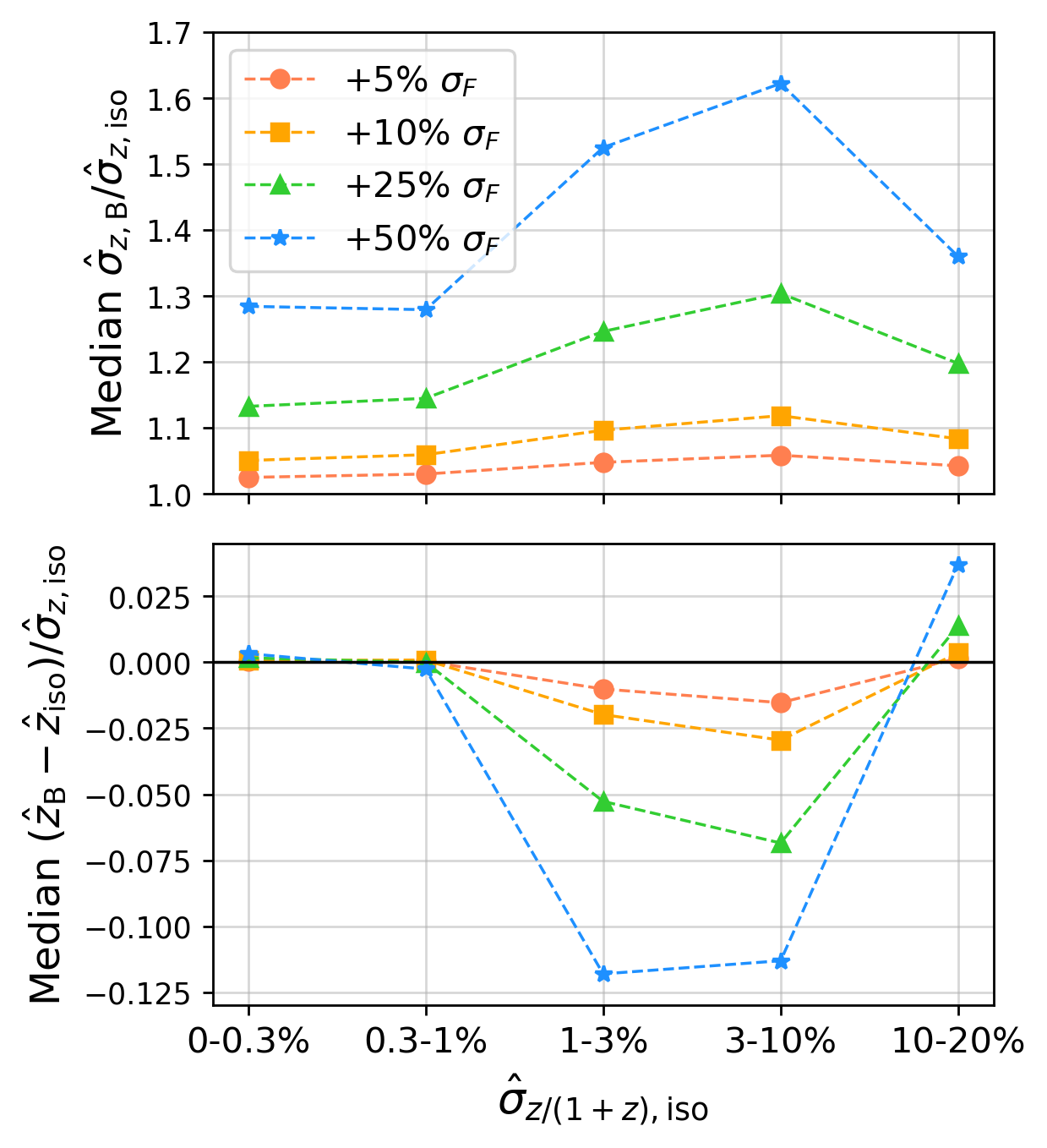}
        \caption{Comparison of blending effects on redshift for the cosmology sample under full-sky sensitivity. We uniformly inflate all galaxy flux errors by $5\%$, $10\%$, $25\%$, and $50\%$, keeping the noise realization fixed to allow consistent comparison. The sample combines both GAMA and COSMOS galaxies to cover the full range of target magnitudes. The \textbf{top} panel shows the fractional increase in redshift uncertainties relative to the isolated case, and the \textbf{bottom} panel shows the additional redshift bias induced by blending, normalized by the isolated redshift precision, $\hat{\sigma}_{z, \rm iso}$, plotted against the isolated redshift errors. }
        \label{fig:sec5.1_zerr}
    \end{figure}

    As shown in Equation~\ref{eq:fluxerr_rise}, the fractional increase in flux uncertainty due to blending depends primarily on the PSF shape and the degree of source overlap, and is largely independent of the per-pixel noise. As a result, the impact of blending on photometry and redshift accuracy is approximately independent of survey depth. In this section, we present results for the full-sky case to illustrate general trends, as the deep field yields similar redshift performance under blending alone. Deep field results are revisited later in Section~\ref{sec5_comb.3}, where they are combined with spectral confusion to highlight differences in survey depth under more complex scenarios. 
    
    \begin{figure}
        \centering
        \includegraphics[width=1.0\linewidth]{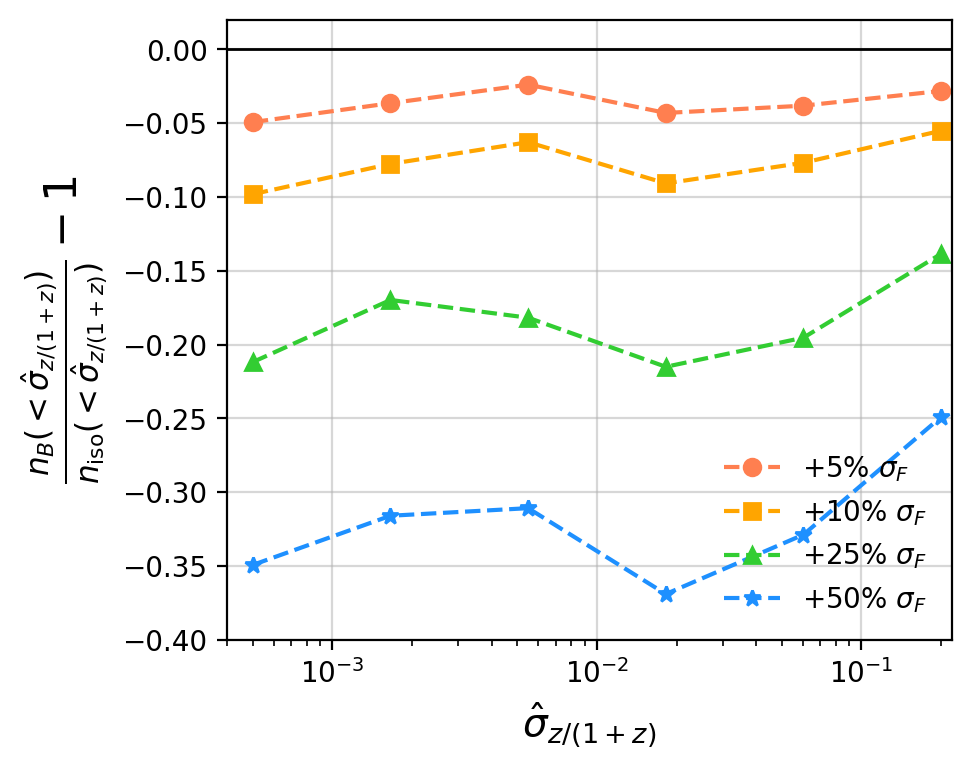}
        \caption{Fractional change in galaxy number density due to blending, as a function of redshift uncertainty threshold under full-sky sensitivity. We show the relative difference between blended ($n_B$) and isolated ($n_{\mathrm{iso}}$) cases, expressed as $\frac{n_B(<\hat{\sigma}_{z/(1+z),B})}{n_{\mathrm{iso}}(<\hat{\sigma}_{z/(1+z),\mathrm{iso}})} - 1$, for both COSMOS and GAMA samples. Results are shown for four levels of artificial flux error inflation.}
        \label{fig:sec5/Nloss_blending}
    \end{figure}

    To gain an intuitive sense of the impact of blended photometry on redshift performance, for example, how a given fractional increase in flux errors propagates to redshift uncertainties, we uniformly scale the isolated photometry uncertainties across all wavelengths by 5\%, 10\%, 25\%, and 50\% and re-inject Gaussian noise based on the scaled errors. We then evaluate the resulting redshift performance for these four levels of blending, as shown in Figure~\ref{fig:sec5.1_zerr}.
    The top panel shows the fractional increase in redshift uncertainties for each redshift precision bin. Redshift errors approximately scale linearly with the photometry error inflation, with galaxies in intermediate precision bins exhibiting the largest impact. The bottom panel illustrates the induced redshift bias relative to the isolated case. The most significant changes occur in the 1-3\% and 3-10\% precision bins. However, the bias remains small. For instance, in the case of a 50\% increase in photometry uncertainty, the redshift bias changes by approximately -12\% of $\hat{\sigma}_{z, \rm iso}$ within the 1-3\% precision bin, corresponding to an overall redshift bias of roughly $-0.3\%$. Notably, the cosmology sample lies near the $+10\%\ \sigma_F$ curve, which induces even smaller bias. These results indicate that while blending inflates both photometry and redshift uncertainties, it does not introduce additional bias.

    The degradation in redshift precision from blending reduces the number of galaxies that meet specific precision requirements. In Figure~\ref{fig:sec5/Nloss_blending}, we quantify this reduction in number density relative to the isolated case. Overall, blending degrades redshift precision, moving galaxies from higher to lower precision bins, which is why the largest reduction occurs in the most stringent precision bin.

\subsection{Spectral Confusion}
\label{sec5:res_S}

        \begin{figure}
            \centering
            \includegraphics[width=1.0\linewidth]{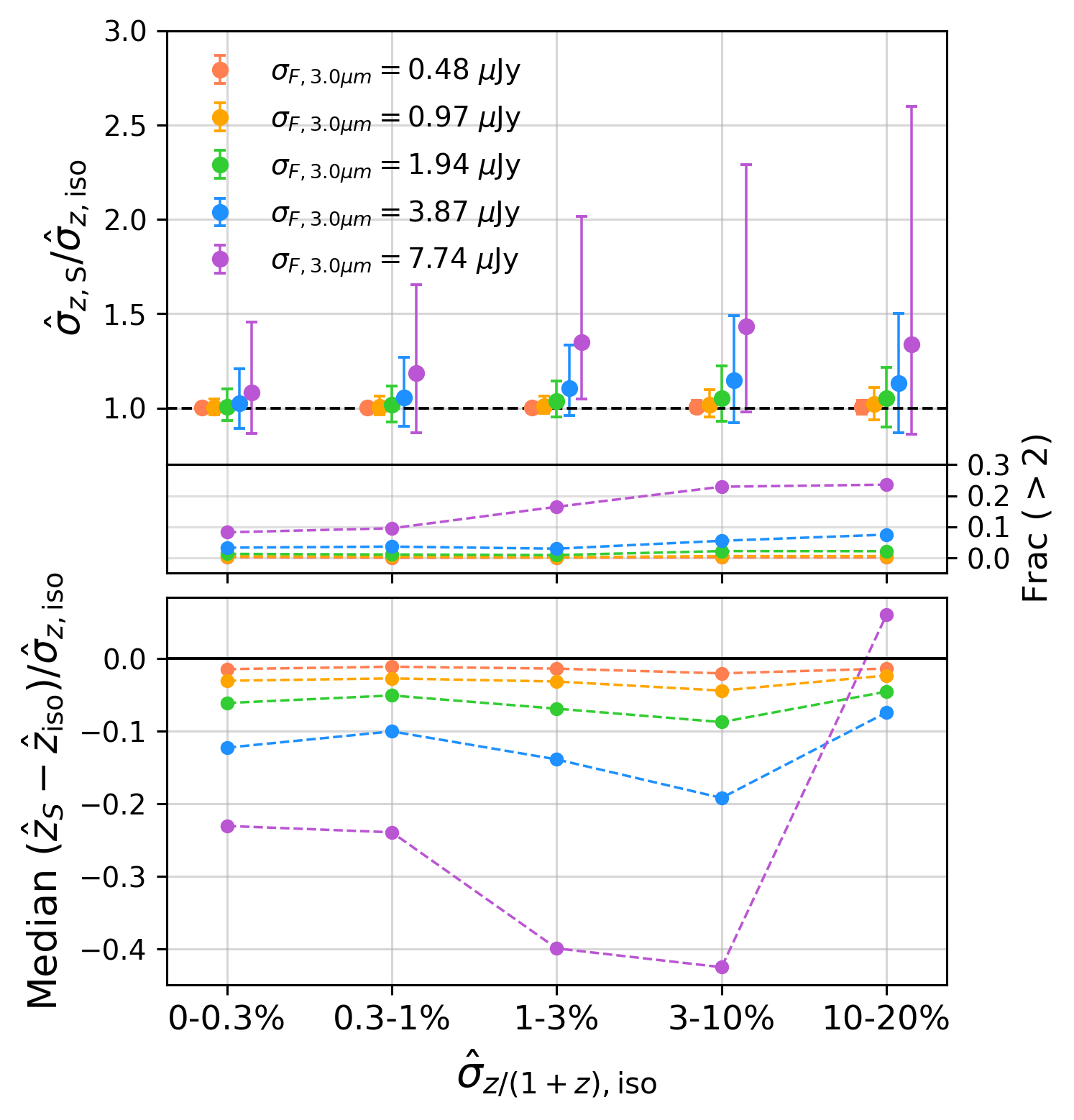}
            \caption{Comparison of spectral confusion effects on redshift for the cosmology sample under full-sky sensitivity. We uniformly scale the confusion library with the cosmology cut shown in Figure~\ref{fig:sec4/lib_variation} and inject them into the cosmology galaxy sample. The library IPR at $3\ \mu\rm m$ is indicated in the legend, with orange error bars showing the unscaled cosmology confusion library ($\sigma_{F, 3.0\mu\rm m}=0.97\ \mu\rm Jy$).
            The \textbf{top} panel shows the fractional increase in redshift uncertainties relative to the isolated case, with error bars marking the 16th, 50th, and 84th percentiles of each distribution. The \textbf{middle} panel displays the fraction of the cosmology sample with $\hat{\sigma}_{z, \rm S} / \hat{\sigma}_{z, \rm iso}>2$. The \textbf{bottom} panel shows the confusion-induced redshift bias.}
            \label{fig:sec5.2a/sigmaZ}
        \end{figure}

        To gain an intuitive understanding of how spectral confusion propagates from photometry to redshift performance, we do a toy experiment and set aside the $z$-magnitude cuts for now. Specifically, we take the confusion library constructed with the fiducial cosmology cut and scale it by factors of 0.5, 1.0, 2.0, 4.0, and 8.0, injecting each set into the cosmology galaxy sample. Similar to the blending test, we examine the resulting redshift error inflation along with the fraction of outliers, and the additional redshift bias, as shown in Figure~\ref{fig:sec5.2a/sigmaZ}. For low confusion levels, the median redshift uncertainties and outlier fraction converge to the isolated case. As the confusion variance increases, both the redshift errors and the outlier fraction rise sharply, with the distribution displaying a pronounced high error tail. The redshift bias remains small near the fiducial cosmology confusion variance, and grows approximately linearly with the confusion variance.

    We now return to the confusion libraries constructed with the $z$-magnitude cuts to assess the impact on number density. The corresponding fractional loss in galaxy counts is shown in the bottom left panel of Figure~\ref{fig:sec5.2b_Nchange} under full-sky sensitivity. Across all selection depths, the number of galaxies meeting the redshift precision thresholds remains essentially unchanged. Together with the redshift bias and outlier fractions presented in Figure~\ref{fig:sec5.2a/sigmaZ} for the cosmology cut, this demonstrates that, under realistic survey conditions, spectral confusion alone does not significantly degrade cosmology sample completeness. Its impact is subdominant to blending and is unlikely to impose a limiting factor at the current selection depth.

    \subsubsection{Varying Survey Depth}
    \label{sec5.2b}

        We assess how the impact of spectral confusion varies with survey depth by comparing redshift degradation under SPHEREx full-sky and deep-field sensitivities. For each $z$-magnitude cut, we inject spectral confusion spectra into galaxy photometry simulated with both noise levels.

        \begin{figure}
            \centering
            \includegraphics[width=1.0\linewidth]{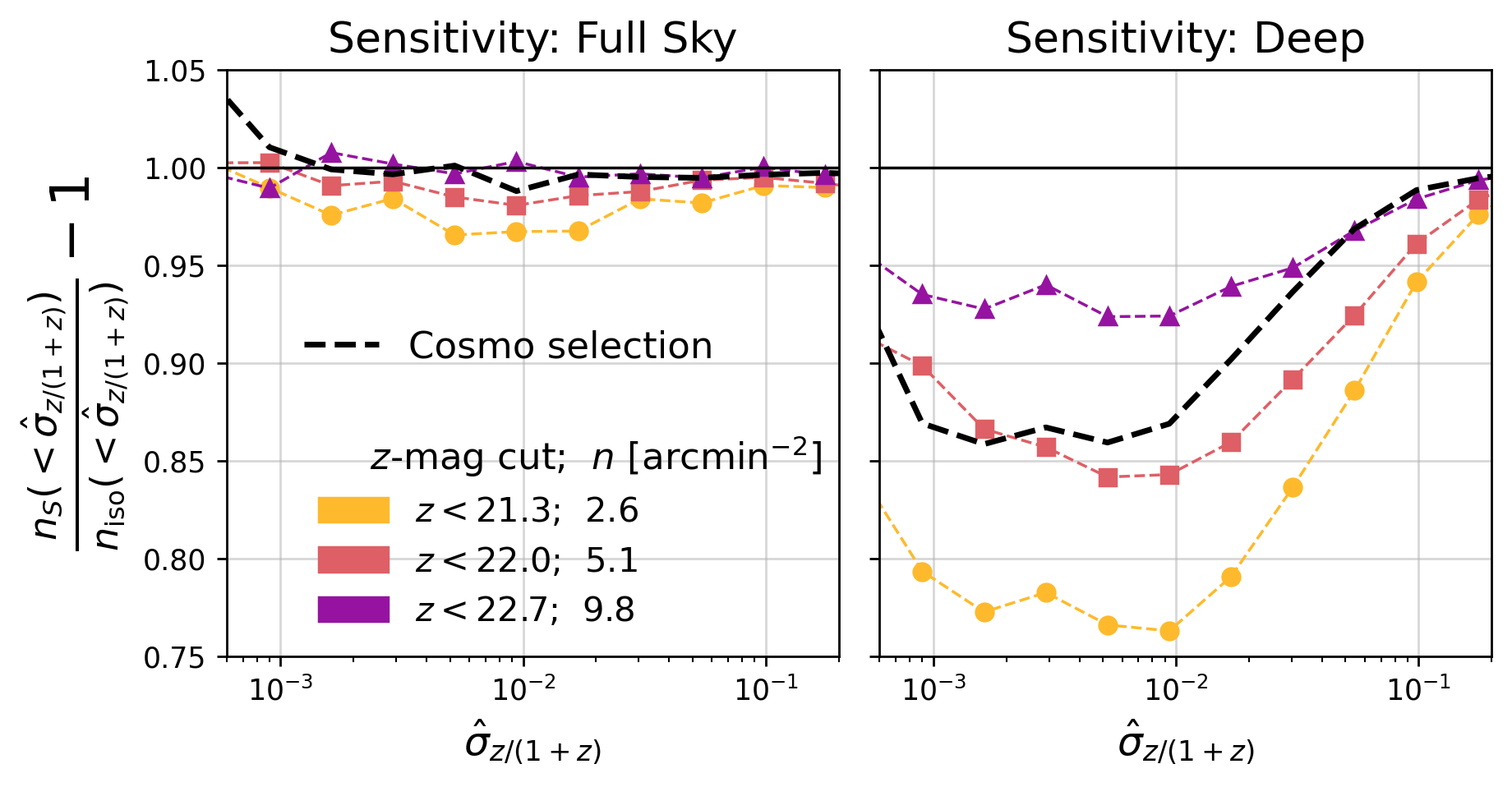}\\[-0.5em] 
            \includegraphics[width=1.0\linewidth]{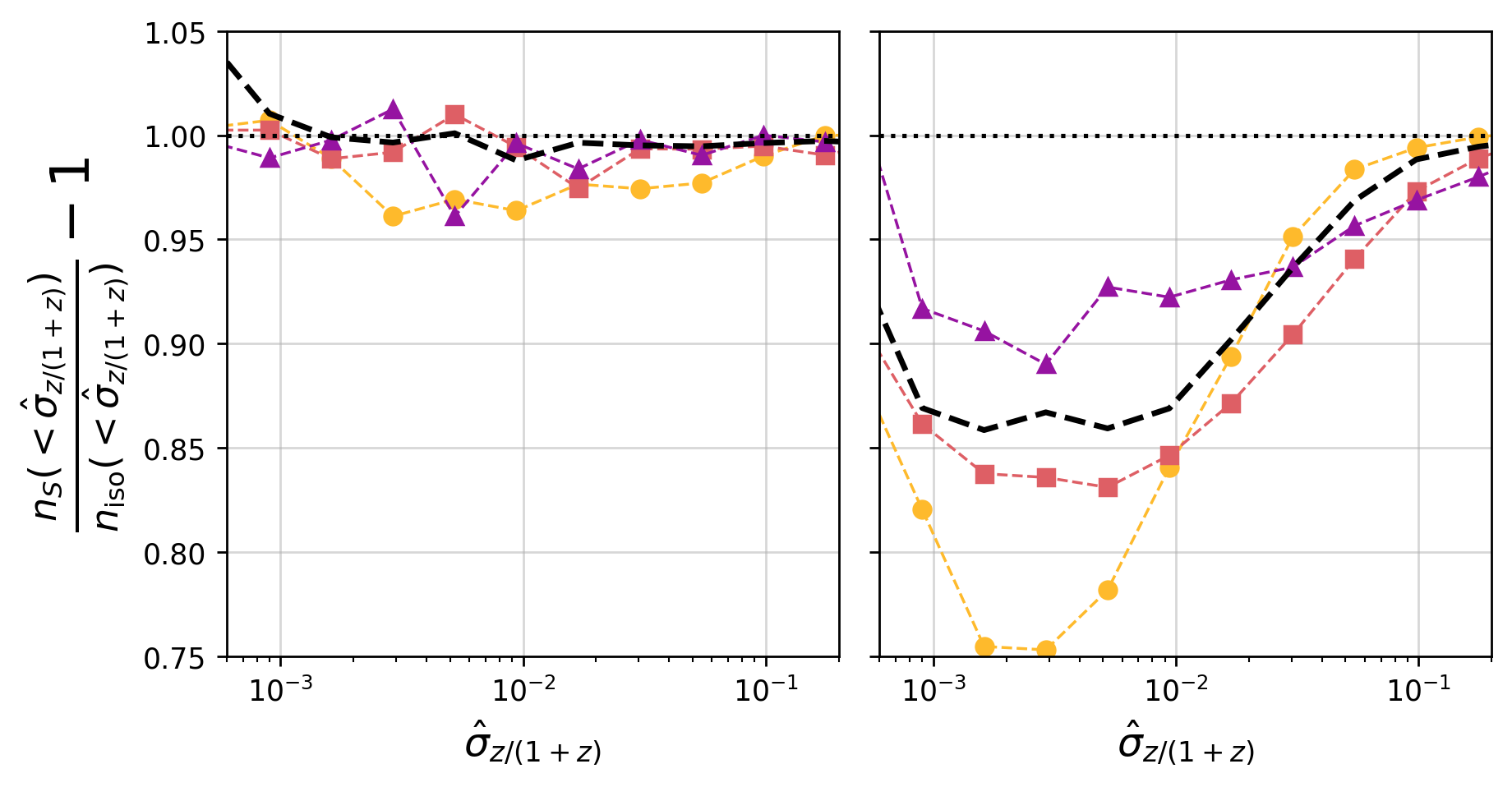}
            \caption{Fractional change in galaxy number density due to spectral confusion as a function of redshift uncertainty threshold, for full-sky (\textbf{left}) and deep-field (\textbf{right}) sensitivity. We show the relative difference between the confusion ($n_\mathrm{S}$) and isolated ($n_\mathrm{iso}$) cases, expressed as
            $\frac{\, n_\mathrm{S}\!\left(<\hat{\sigma}_{z/(1+z),S}\right) \,}{\, n_\mathrm{iso}\!\left(<\hat{\sigma}_{z/(1+z),\mathrm{iso}}\right) \,} - 1$,
            using the combined COSMOS and GAMA samples. Only the first three $z$-magnitude cuts are shown, as these are most relevant for the cosmology study; the faintest cut is omitted. The \textbf{top row} shows results for the same galaxy sample (cosmology selection) with confusion libraries from different $z_{\rm mag}$ cuts, enabling a consistent comparison. The \textbf{bottom row} shows the realistic case where the galaxy sample and the confusion library correspond to each $z$-magnitude cut; the crossovers in the deep-field curves arise from differences in selected galaxies. Results from the cosmology-selection confusion library are highlighted with black dashed lines.}
            \label{fig:sec5.2b_Nchange}
        \end{figure}

        Figure~\ref{fig:sec5.2b_Nchange} shows the fractional number loss due to confusion alone, relative to the isolated case. The top panels use different confusion libraries from $z$-magnitude cuts injected into the same cosmology sample for consistency. The bottom panels show the realistic case where both galaxies and confusion are selected by each $z$-magnitude cut, and the resulting crossovers in the deep-field curves reflect differences in the selected galaxies. 

        Compared to the full-sky case, the deep field shows a more pronounced fractional loss, though it still delivers a much higher overall redshift density. Interestingly, the impact is not monotonic: the loss is relatively small at the very highest-precision end, grows most severe in the intermediate-precision bins, and then diminishes again toward the lowest-precision end. This trend indicates that spectral confusion primarily degrades intermediate-precision galaxies, pushing them into lower-precision bins, while the very brightest galaxies remain resilient and the faintest are already dominated by other noise sources. 
        
        In the realistic case (bottom right panel), the crossovers between different $z$-magnitude can be explained as follows. Shallow cuts select brighter galaxies with higher native precision, so although confusion degrades some of them, their performance remains relatively strong and the loss closes quickly. Deeper cuts target fainter galaxies with poorer baseline precision, and confusion has a smaller relative impact. But once degraded, these galaxies cannot recover, leading to more persistent losses.

        \begin{figure}
            \centering
            \includegraphics[width=1.0\linewidth]{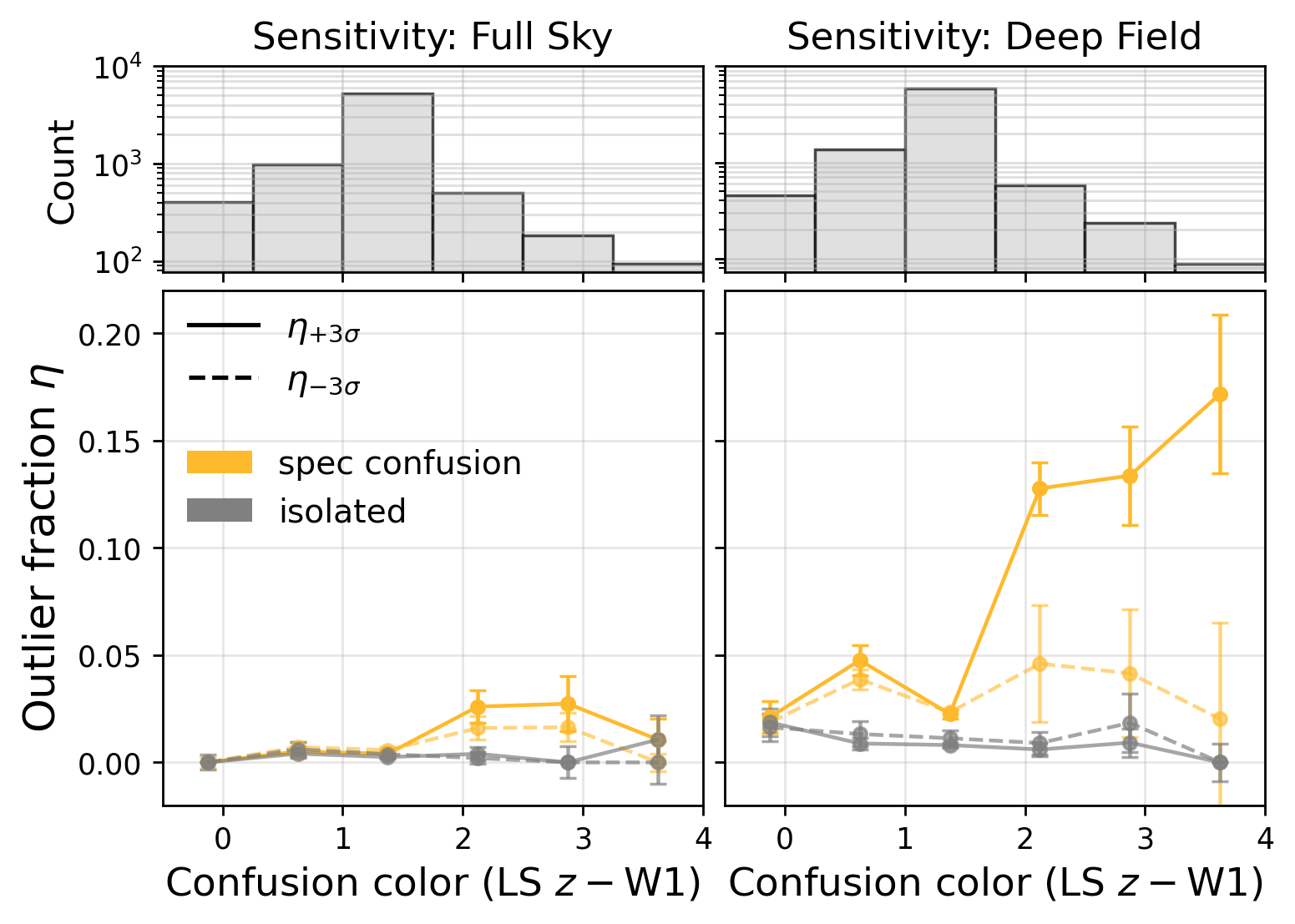}
            \caption{Increase in redshift outlier fraction due to spectral confusion, shown as a function of confusion color for sources with the $z_{\rm mag}<21.3$ selection cut and $\hat{\sigma}_{z/(1+z),\mathrm{iso}} < 0.2$. The bottom panels compare the isolated case (grey) and spectral confusion case (yellow) under full-sky (left) and deep-field (right) sensitivities. Overestimated (solid) and underestimated (dashed) $3\sigma$ outliers --- defined as $\frac{\hat{z} - z_{\mathrm{true}}}{1 + z_{\mathrm{true}}} > 3\hat{\sigma}_{z/(1+z)}$ and $< -3\hat{\sigma}_{z/(1+z)}$, respectively --- are denoted by $\eta_{+3\sigma}$ and $\eta_{-3\sigma}$. We compute confusion color by convolving confusion spectra with LS-$z$ and WISE-1 filters. We find a strong correlation between confusion color and the excess of overestimated outliers, while the underestimated outlier fraction shows only weak dependence. The top panel shows the distribution of sources in confusion color, revealing that the large outlier fractions occur for only a small subset of the sample.}
            \label{fig:sec5.2b_bias}
        \end{figure}

        Beyond redshift precision, we examine redshift outliers, since spectral confusion can introduce correlated spectral features. Figure~\ref{fig:sec5.2b_bias} shows the $3\sigma$ outlier fractions for both overestimated and underestimated redshifts as a function of the LS-$z-\mathrm{W1}$ color of each injected confusion spectrum, using the $z$-magnitude cut at 21.3 as a representative case. In the confusion-free simulation, outlier fractions are flat with no dependence on confusion color. In contrast, confusion-injected samples, especially under deep-field sensitivity, show a strong positive correlation between confusion color and overestimated redshift outliers, indicating a systematic bias toward higher redshifts. The trend is consistent with the red excess in the spectral confusion library (Figure~\ref{fig:sec4/confusion_lib}), driven by both PSF broadening at longer wavelengths and the prevalence of redder background sources. This cannot be captured by continuum fitting $\chi^2$ or other metrics. While the overestimated outlier fraction reaches up to 17\% at the reddest confusion colors, the number density of such extreme confusion spectra is small. Most confusion colors lie in the range $z$-$\mathrm{W1} \sim$ 0–1 mag, where outlier fractions are much lower. As a result, the total statistical impact on the overall sample’s outlier rate remains modest. 
        
        In contrast, the underestimated outlier fraction shows no clear correlation with confusion color and remains statistically consistent with the isolated case across the color range. This asymmetry reinforces the conclusion that spectral confusion tends to bias redshifts high, rather than producing symmetric scatter.

        For the current cosmology sample after including spectral confusion, the outlier fraction changes from $0.8\%$ to $1.3\%$ under full-sky sensitivity, and from $2.3\%$ to $4.5\%$ in the deep field. These results imply that while the average impact of spectral confusion on redshift outliers is minimal, systematic high-redshift biases can arise in specific regimes, especially in deep surveys. 

\subsection{Combining Blending and Spectral Confusion}
\label{sec5_comb.3}

    \begin{figure*}
        \centering
        \includegraphics[width=1.0\linewidth]{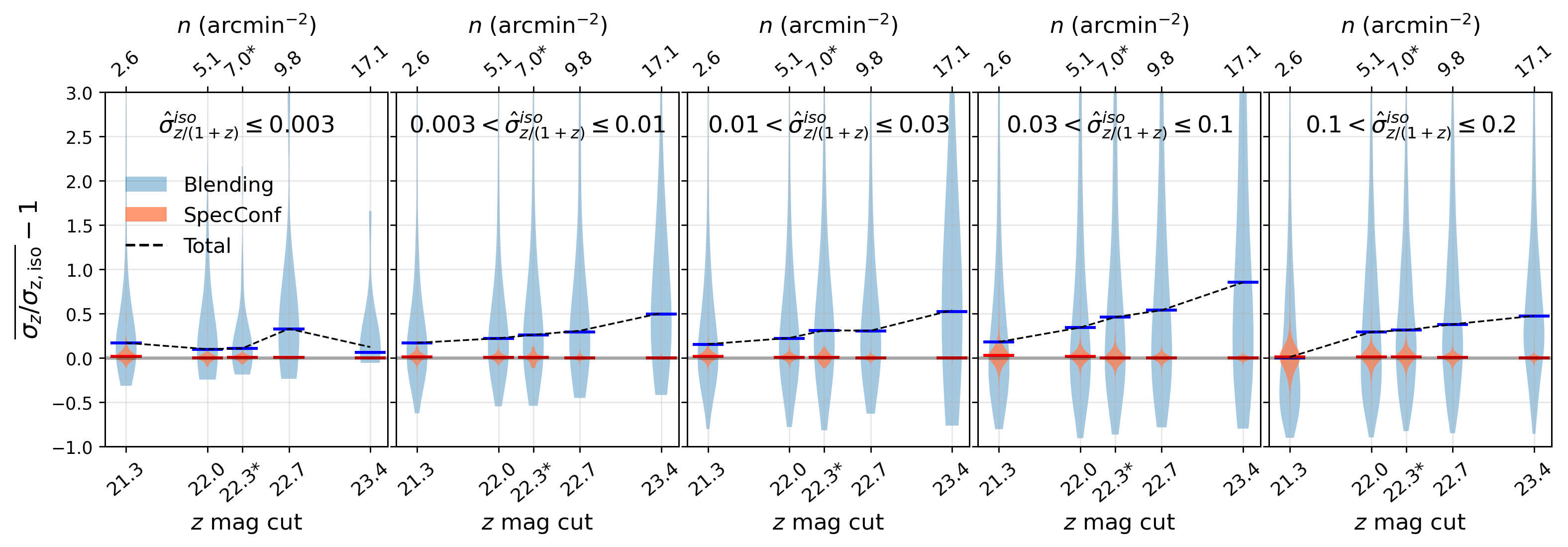}
        \caption{Fractional increase in redshift uncertainty from blending and spectral confusion under full-sky sensitivity, as a function of target density. Blue violins represent the effect of blending (median marked), and orange violins show the effect of spectral confusion (median marked). Dashed black lines represent the quadrature sum of both effects, assuming they are independent; in practice, they may couple nonlinearly. 
        Target density is labeled on the upper x-axis, with corresponding $z$-mag cuts shown on the lower x-axis. Each violin corresponds to a galaxy sample selected by the indicated $z$-magnitude cut. The violins marked with an asterisk ($z<22.3$) represent the fiducial cosmology sample, plotted at 22.3 for visualization since it yields a similar target density to that cut. Results are binned by isolated redshift uncertainty, $\hat{\sigma}^{\mathrm{iso}}_{z/(1+z)}$. }
        \label{fig:sec5/BS/sigz_acc_full}
    \end{figure*}

    We compare blending and spectral confusion, and assess the combined impact on redshift performance. We begin by presenting their individual contributions side-by-side across different selection depths and survey sensitivities, and then evaluate the total degradation when both are applied simultaneously. This analysis helps identify potential optimal selection depths where redshift degradation is minimized, informing sample optimization for cosmological analyses.

    Under full-sky sensitivity, Figure~\ref{fig:sec5/BS/sigz_acc_full} shows the fractional increase in redshift uncertainty relative to the isolated case, displaying the full distribution using violin plots. This expands on the trends seen in Figures~\ref{fig:sec5.1_zerr} and \ref{fig:sec5.2a/sigmaZ}. We also include a naive quadrature sum of the individual blending and spectral confusion effects, assuming that they are independent. In practice, they may couple nonlinearly.
    Overall, blending clearly dominates the total degradation. Its impact increases at deeper selection cuts due to higher target density. The redshift uncertainty inflation is smallest for the highest precision bin, then grows across bins of decreasing precision, until flattening at the lowest-precision end.

\begin{figure*}
        \centering
        \includegraphics[width=1.0\linewidth]{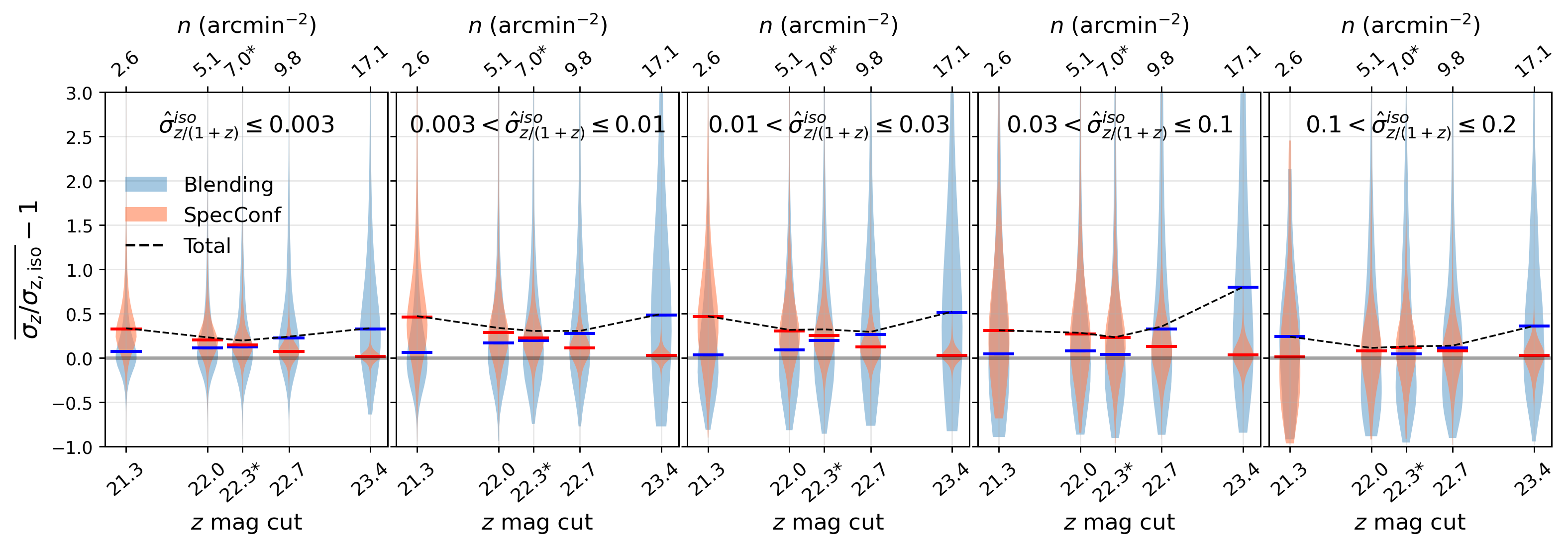}
        \caption{Same as Figure~\ref{fig:sec5/BS/sigz_acc_full}, but showing results under the deep field sensitivity. Note that in the last, lowest precision bin, the violin distributions at the $z_{\rm mag}<21.3$ cut are slightly off the trend both in blending and spectral confusion. This can be explained by the small sample of galaxies falling into this low precision bin given the bright cut. }
        \label{fig:sec5/BS/sigz_acc_deep}
    \end{figure*}

    Under deep-field sensitivity, Figure~\ref{fig:sec5/BS/sigz_acc_deep} shows that spectral confusion becomes comparable to blending in driving redshift degradation. In contrast, blending remains largely unchanged from the full-sky case and does not depend on survey depth. This is consistent with Equation~\ref{eq:fluxerr_rise}, where photometric degradation due to blending scales primarily with source separation and PSF size. We observe a clear trade-off: within each precision bin, spectral confusion decreases with deeper cuts as fewer untargeted sources remain to contribute excess flux, while blending increases due to the higher surface density of targets. The quadrature sum highlights this tension and shows an optimal spot in selection depth that minimizes the combined effects. 

\begin{figure}
        \centering
        \includegraphics[width=1.0\linewidth]{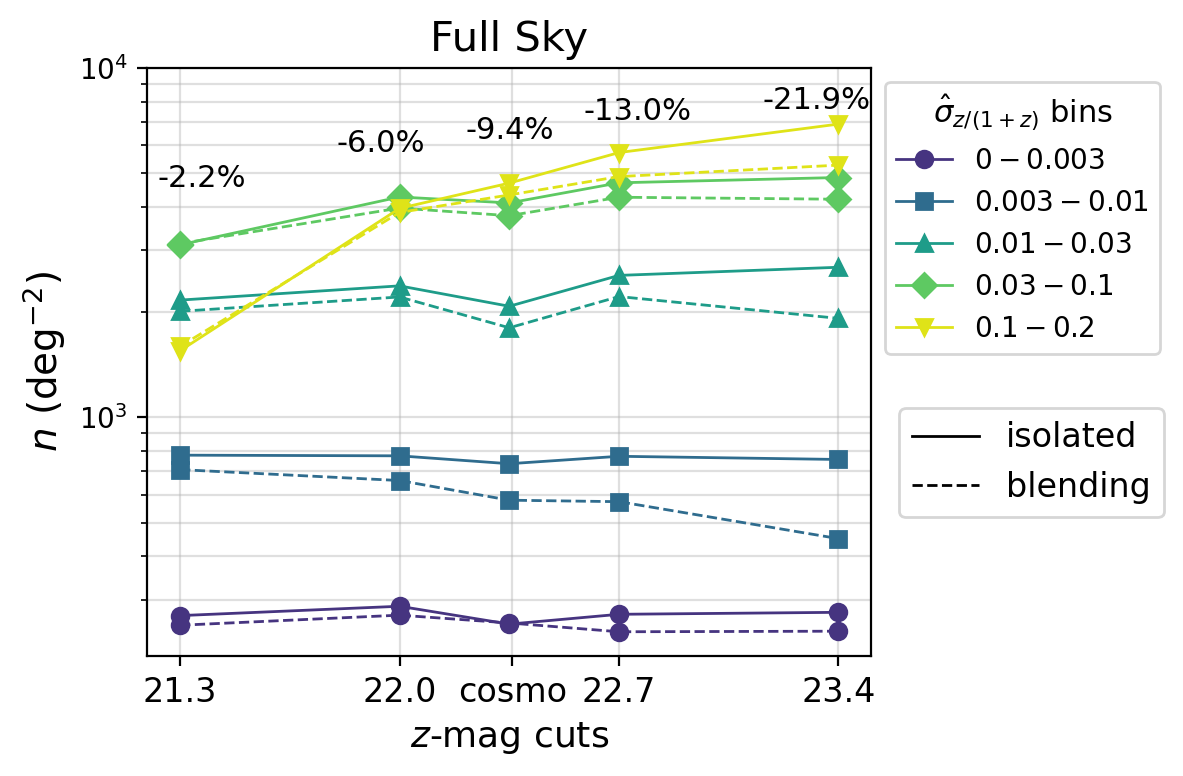}
        \caption{Total galaxy number density under full-sky sensitivity, combining COSMOS and GAMA samples, as a function of selection depth (the cosmology sample is marked as `cosmo' on the x-axis). Only the impact of blending is included, since spectral confusion is negligible at this depth. For each redshift precision bin $\hat{\sigma}_{z/(1+z)}$, solid lines show the isolated number density estimate, while dashed lines show the blending-impacted case. Numbers above the curves indicate the total fractional loss for $\hat{\sigma}_{z/(1+z)}<0.2$ given each selection cut. Number densities are computed using SPHEREx-only 102 bands, without any external photometry.}
        \label{fig:sec5/Nloss_combined_full}
    \end{figure}

    Complementing this, Figure~\ref{fig:sec5/Nloss_combined_full} shows the resulting fractional loss in redshift samples under full-sky sensitivity. Here, blending is the dominant contributor and thus spectral confusion is neglected. We observe a consistent decrease in number density due to blending, even when extending to deeper selections, particularly in the $0.003<\hat{\sigma}_{z/(1+z)} < 0.01$ precision bin. 

    \begin{figure*}[!ht]
        \centering
        \includegraphics[width=1.0\linewidth]{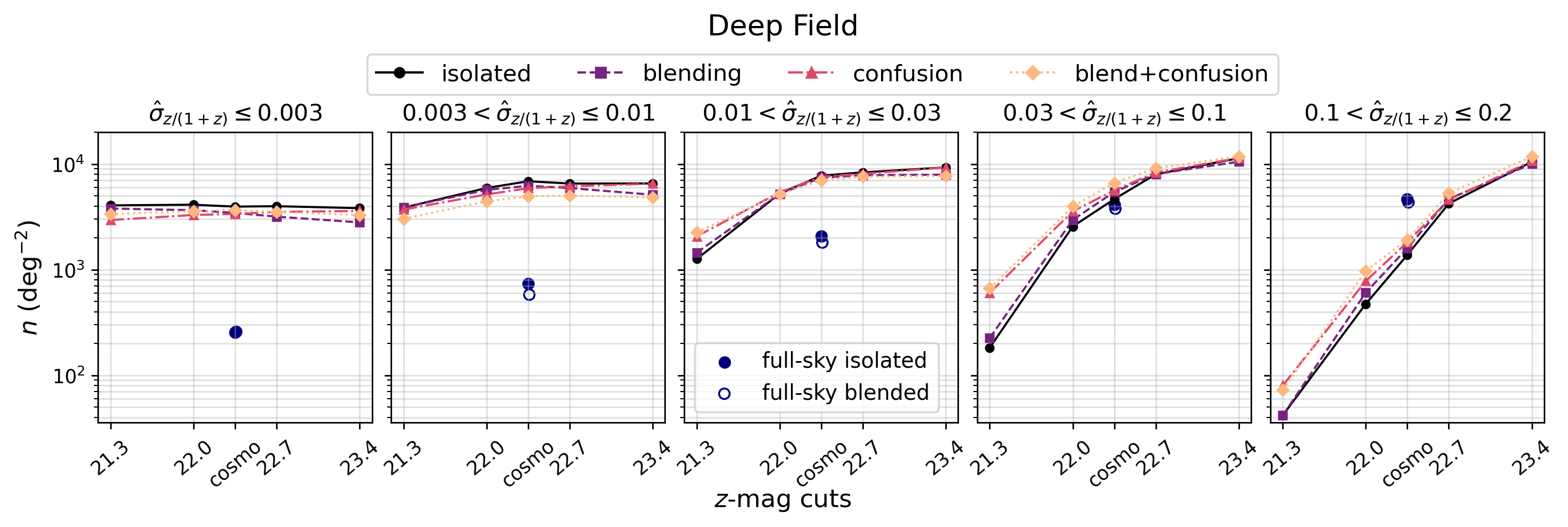}
        \caption{Total galaxy number density under deep-field sensitivity. For each redshift precision bin $\hat{\sigma}_{z/(1+z)}$, we show four cases: isolated (solid), blending only (dashed), spectral confusion only (dash-dotted), and blending plus confusion (dotted). The full-sky cosmology number density is marked with navy circles. Fractional losses of the $\hat{\sigma}_{z/(1+z)}<0.2$ sample for each $z$-band cut are: 21.3: 0.39\%, 22.0: 1.00\%, cosmo: 1.73\%, 22.7: 1.69\%, 23.4: 5.10\%. 
        Number densities are computed using SPHEREx-only 102 bands, without any external photometry.}
        \label{fig:sec5/Nloss_combined_deep}
    \end{figure*}

    We also present the corresponding number density changes under deep-field sensitivity in Figure~\ref{fig:sec5/Nloss_combined_deep}, comparing the effects of blending only, confusion only, and their combination. 
    In the isolated deep-field case, the highest-precision bins gain substantially relative to the full-sky. As a result, the lowest-precision bins show reduced number densities because many sources are promoted to higher-precision bins, and our catalog depth limits the availability of additional faint sources in the deep field. 

    Blending and spectral confusion redistribute some high-precision sources into lower-precision bins, producing net gains there relative to the isolated case. Overall, the combined effect leads to higher total number densities at deeper cuts, particularly in the lower-precision bins. Notably, the LS-$z$ magnitude cut that minimizes the fractional impact of confusion and blending in Figure~\ref{fig:sec5/BS/sigz_acc_deep} does not necessarily maximize the absolute redshift density.

\section{Discussion}
\label{sec:discussion}

Our simulations have so far been based exclusively on COSMOS and GAMA galaxy samples. However, stellar contamination remains a significant factor that must be considered in the context of source blending and spectral confusion. We also discuss potential optimization of deep-field selection to maximize scientific outcomes.

\subsection{Stellar Contamination}

To assess stellar contamination under our fiducial cosmology selection, we use TRILEGAL \citep{trilegal05}, a stellar population synthesis code that models the photometric properties of Milky Way stars based on stellar evolution and Galactic structure. We simulate stellar populations at Galactic longitudes $l=90^\circ$, $180^\circ$, and $270^\circ$, across latitudes $|b|>25^\circ$. The stellar fraction among selected targets decreases with latitude, from $\sim47\%$ at $b=25^\circ$ to $\sim12\%$ near the Galactic poles ($|b|>85^\circ$), with a mean of $\sim22\%$. In terms of selected stellar density, this corresponds to $n\sim5.6$ arcmin$^{-2}$ at $b=25^\circ$ and $n\sim1.2$ arcmin$^{-2}$ near the poles. Stellar density also declines more rapidly than galaxy density toward fainter magnitudes, contributing only $\sim4\%$ of sources near the cosmology selection threshold.

Although stars contribute minimally to spectral confusion due to their low number density at faint magnitudes, their combined flux is relatively stronger at shorter wavelengths and may partially counteract the redder confusion from extragalactic sources. At brighter magnitudes, however, their comparatively higher density and their extended profiles make them a significant source of blending. We estimate the fractional loss in effective number density due to stellar blending using Figure~\ref{fig:sec5/Nloss_blending}. To first order, stars increase the total targeted source density, which is the dominant factor driving blending. Assuming an average targeted stellar fraction of $\sim\!22\%$, the total target density after the cosmology cut increases from $7.0\ \mathrm{arcmin}^{-2}$ to approximately $8.5\ \mathrm{arcmin}^{-2}$, increasing the loss from $9.4\%$ in Figure~\ref{fig:sec5/Nloss_combined_full} to just under 13\% for the $\hat{\sigma}_{z/(1+z)}<0.2$ sample. A more complete treatment of stellar blending is left to future work, as it involves multiple interdependent factors such as source masking, PSF reconstruction, and local source environments. Star-galaxy classification errors can further complicate both confusion and blending.

\subsection{Deep Field Optimization}

    SPHEREx’s survey strategy yields much better sensitivity in deep fields, enabling more accurate redshift estimates and allowing for deeper selection cuts. The deep survey supports enhanced studies of galaxy formation and line intensity mapping. 
    
    Under deep-field sensitivity, we find a trade-off: while spectral confusion decreases with more targeted sources, blending increases. Modestly deeper cuts can still provide a net gain in the number density of reliable redshifts in the deep field, despite increased blending, as indicated by Figure~\ref{fig:sec5/Nloss_combined_deep}. This supports our deep-field selection campaign to improve galaxy cross-correlation S/N \citep{YTC22}.
    
    In addition, the higher incidence of overestimated redshift outliers motivates adopting a redder selection cut, i.e., extending deeper in the W1 band to flatten the spectral slope of confusion spectra (Figure~\ref{fig:sec4/confusion_lib}) and thereby mitigate redshift bias.

\section{Conclusion}
\label{sec:conclusion}

In this work, we present methods to model two key effects that will impact SPHEREx photometry and redshift estimation: blending from overlapping targeted sources, and spectral confusion from integrated flux of untargeted background galaxies. Both contribute to degraded spectral fidelity, although through different mechanisms --- blending amplifies photometric noise while spectral confusion induces systematic color biases.

For blending, we find that flux measurements remain unbiased, but uncertainties increase with target density and source proximity. As the selection depth increases, resulting in higher target densities, blending effects become more pronounced across all redshift precision bins. Given the current cosmology selection, blending removes about 9.4\% of galaxies from the $\hat{\sigma}_{z/(1+z)}<0.2$ sample.

To model spectral confusion, we develop a Monte Carlo method using realistic background galaxy SEDs, capturing the spectral variance beyond traditional confusion noise estimates used in source detection. While confusion has negligible effect at full-sky depth, it becomes significant at deep-field depth. At shallow selection limits in the deep field, spectral confusion dominates over blending in redshift degradation, but its influence decreases with depth, eventually falling below that of blending. We also find that confusion introduces a color bias -- redder confusion spectra systematically bias galaxy redshifts high, leading to overestimated redshifts and increased rates of outliers.  

When considering both blending and spectral confusion together, we can further optimize the deep-field selection to balance these competing effects, maximizing the overall redshift density. 

However, additional spectral features may introduce further complications. In particular, strong emission lines in star-forming galaxies may contribute to confusion and blending. They may lead to systematic errors in redshift estimates or derived galaxy properties. This effect has not been explored in the current work and remains a subject for future study.

Since both blending and spectral confusion are coupled to the source density field, they will also introduce correlated redshift errors. Future work will focus on generating systematics templates that can be fitted and removed from observations to mitigate this density-coupled effect. While correlations induced by the galaxy density field are confined to small angular scales and are not expected to impact $f_{\mathrm{NL}}$ constraints, variations in the stellar density can produce larger-scale systematics that will require more careful treatment in future analyses.



\section*{Acknowledgments}
We acknowledge support from the SPHEREx project under a contract from the NASA/Goddard Space Flight Center to the California Institute of Technology. Part of the research described in this paper was carried out at the Jet Propulsion Laboratory, California Institute of Technology, under a contract with the National Aeronautics and Space Administration (80NM0018D0004). Y. K. was supported by the National Research Foundation of Korea (NRF) grant funded by the Korean government (MSIT) (No. 2021R1C1C2091550). Bomee Lee was supported by the National Research Foundation of Korea (NRF) grant funded by the Korean government(MSIT) (No. 2022R1C1C1008695).


%



\bibliographystyle{aasjournalv7} 
\bibliography{main}

\appendix

\section{Derivation of Flux Covariance}
\label{appendix:secA}

        We present the derivation of flux covariance between overlapping targets. For a point source $i$ with image coordinates $(x_i,y_i)$ and flux $f_i$, the corresponding pixelized intensity distribution is given by 
        \begin{equation}
            I(x,y) = f_i p(x-x_i, y-y_i).
            \label{eq:Ixy}
        \end{equation}
        For simplicity, we denote the pixel-convolved PSF as $p_i(x,y)$. 
        
        For Gaussian free parameter $\theta$, we write down the log-likelihood, 
        \begin{equation}
            \text{ln}\mathcal{L}(\hat{\boldsymbol{f}}|\theta) = -\frac{N_\mathrm{pix}}{2}\text{ln}(2\pi \sigma^2) -\frac1{2} \sum_{x,y}^{N_\mathrm{pix}} \left(\frac{\hat{f}(x,y) - \sum_i^{N_{src}}f_ip_i(x, y)}{\sigma(x,y)}\right)^2,
        \end{equation}
        where $N_{\mathrm{pix}}$ is the total number of pixels. 
        We obtain the best-fit flux estimates by maximizing the likelihood, i.e., by setting the first derivative of the likelihood with respect to flux to zero. The corresponding covariance matrix is given by the inverse FIM, computed from the second derivatives: 
        \begin{equation}
            \begin{split}
                \mathrm{cov}&(\theta_{\mathrm{ML}}) = \mathcal{F}^{-1}(\theta_{\mathrm{ML}}) \\
                &= -\begin{bmatrix}
                    \partial^2_{f_1} \mathrm{ln}\mathcal{L}(\theta_{\text{ML}}) & \partial_{f_1}\partial_{f_2}\mathrm{ln}\mathcal{L} & \cdots & \partial_{f_1}\partial_{f_k}\mathrm{ln}\mathcal{L} \\
                    \partial_{f_2}\partial_{f_1}\mathrm{ln}\mathcal{L} & \partial^2_{f_2} \mathrm{ln}\mathcal{L} & \cdots & \partial_{f_2}\partial_{f_k}\mathrm{ln}\mathcal{L} \\ 
                    \vdots & \vdots & \ddots & \vdots \\
                    \partial_{f_k}\partial_{f_1}\mathrm{ln}\mathcal{L} & \partial_{f_k}\partial_{f_2}\mathrm{ln}\mathcal{L} & \cdots & \partial_{f_k}^2\mathrm{ln}\mathcal{L}
                \end{bmatrix}^{-1}_{\theta=\theta_{\mathrm{ML}}}
            \end{split}
            \label{eq:FIM}
        \end{equation}
        for a $k$-source Tractor fit $\theta_{\mathrm{ML}}=\{f^{\mathrm{ML}}_i\}_{i=1}^k$. 
        We specify the FIM elements evaluated at the maximum-likelihood parameters. Diagonal terms correspond to individual sources, while off-diagonal terms capture the covariance between overlapping source pairs. These are given by
        \begin{equation}
            \partial^2_{f_a}\mathrm{ln}\mathcal{L} = \sum_{x,y}^{N_{\mathrm{pix}}} \frac{p_a^2(x, y)}{\sigma^2(x,y)},
            \label{eq:partial2}
        \end{equation}
        \begin{equation}
            \partial_{f_b}\partial_{f_a}\mathrm{ln}\mathcal{L} = \sum_{x,y}^{N_\mathrm{pix}} \frac{p_a(x, y)p_b(x, y)}{\sigma(x,y)^2},
            \label{eq:crosspartial}
        \end{equation}
        where $a$ and $b$ index any pair of sources in the fit. This is numerically implemented to account for blending-induced flux covariance. We do not analytically expand the inverse FIM beyond the two-source case in this work.

\end{CJK*}
\end{document}